# Enhanced thermo-mechanical properties of 212 MAX phase borides $Zr_2AB_2$ (A = In, Tl): an ab-initio understanding

M. A. Ali[a,*], M. M. Hossain[a], M. M. Uddin[a], A. K. M. A. Islam[b,c], S. H. Naqib[b,**]

[a]Department of Physics, Chittagong University of Engineering and Technology (CUET), Chattogram 4349, Bangladesh
[b]Department of Physics, University of Rajshahi, Rajshahi-6205, Bangladesh
[c]Department of Electrical and Electronic Engineering, International Islamic University Chittagong, Kumira, Chattogram, 4318, Bangladesh

**Abstract**

The discovery of MAX phase borides has added a new dimension for research in the materials science community. In this paper, a first-principles study of the newly known MAX phase borides $Zr_2AB_2$ (A = In, Tl) has been carried out. The stiffness constants, elastic moduli, mechanical anisotropy, thermal and optical properties are investigated for the first time. The estimated lattice constants and volumes of the unit cell are found to be consistent with previous study. The dynamical and mechanical stability of the titled compounds have been checked adroitly by calculating the phonon dispersion curves and stiffness constants, respectively. Fundamental insights into the stiffness constants, elastic moduli, hardness parameters, brittleness and anisotropy indices are presented. The variation of these mechanical properties is explained based on the Mulliken population analysis and charge density mapping (CDM). The electronic properties have been dealt with by considering electronic band structure and density of states (DOS) which confirmed the metallic nature of $Zr_2AB_2$ (A = In, Tl). The lowly dispersive energy bands along the *c*-direction confirmed anisotropy in conductivity. The DOSs are almost identical for both the compounds and are consistent with the results obtained from CDMs. The analysis of DOS revealed the dominant contribution from Zr-d orbitals to the conductivity with a small contribution from the In/Tl-*p* states contributing at the Fermi level. Important thermal properties such as Debye temperature ($\Theta_D$), minimum thermal conductivity ($K_{min}$), Grüneisen parameter ($\gamma$) and melting temperature ($T_m$) have been calculated. The higher values of $\Theta_D$ and $T_m$, and lower value of $K_{min}$ for $Zr_2AB_2$ (A = In, Tl) compared to those of $Zr_2AC$ (A = In, Tl) suggest enhanced thermal properties of the compounds under study for practical applications. Besides, the specific heat capacities ($C_v$, $C_p$), thermal expansion coefficient, and different thermodynamic potential functions have been calculated from the phonon density of states. The technologically important optical constants have been studied with an intention to reveal their possible relevance for application purposes. The reflectivity spectra revealed the applicability of $Zr_2AB_2$ (A = In, Tl) as cover materials to diminish the solar heating in the IR, visible, and near UV regions. The studied physical properties of $Zr_2AB_2$ (A = In, Tl) are compared with those of other relevant 212 and 211 MAX phase nanolaminates.

Corresponding authors: [*]ashrafphy31@cuet.ac.bd; [**]salehnaqib@yahoo.com

## 1. Introduction

Ever since the MAX phase materials have been brought into light by Barsoum et al. [1–3], these materials have drawn extensive attention by materials scientists and engineers because of their fascinating combination of physical properties [4–13]. The MAX phase materials exhibit metallic as well as ceramic behavior owing to the existence of alternative metallic A-layer and ceramic MX layer. The unconventional bridging of metallic and ceramic properties [14] makes them multifunctional compounds for use in various sectors as (i) an alternative to graphite at high-temperatures, (ii) heating elements, (iii) high temperature foil bearings and other tribological components, gas burner nozzles, and (iv) a tool for dry drilling of concrete [13]. Moreover, these compounds can also be used in sensors, electrical contacts, fuel cells, nuclear industry and spintronics [9,15,16]. A recent important application of MAX phase materials is as the precursor for 2D MXenes. The MXenes are not synthesizable directly owing to their thermodynamical meta-stability [17−19]. The derivation of MXenes from MAX phases is possible due to the existence of weak M-A bonding from where the A layer can be selectively etched.

However, since their discovery, the diversity of the MAX phases has been limited to C and N as X elements for many years. Recently, this limitation has been overcome by the successful synthesis of B containing MAX phases [20–22]. Moreover, the scientific communities are also trying to extend their diversity by changing the structure such as $Mo_2Ga_2C$ [6,23], 321 MAX phases [24], atomically layered and ordered rare-earth i-MAX phases [25,26], 314 MAX phases [27] and the 212 compounds [28]. In the last two groups, 314 MAX phase [27] and 212 MAX phase [28,29], the atom B has been used as an X element, a recent addition to the MAX family. Though one of 212 MAX ($Ti_2InB_2$) phase borides has already been synthesized [29], others are expected to be thermodynamically stable as investigated by Miao *et al.*[30].

Several attempts have already been made to synthesize the MAX phase borides drawing the inspiration from the properties of conventional C and N containing MAX phase materials. The first step was being successful by the discovery of MAB phase materials [31], already able to draw significant attention [32–34]. The MAB phases are somewhat different from the conventional MAX phases because of their structures are crystallized in the orthorhombic symmetry [31,32]. $Ti_2InB_2$, named as 212 MAX phase, was first synthesized by Wang *et al.*[29]

who were inspired by the results reported by Ade *et al.* [31]. The structure is slightly different from the conventional MAX phases which crystallized in the hexagonal system with a space group $P\bar{6}m2$ (No. 187). Later, the structural stability, mechanical properties and thermalconductivity of $Ti_2InB_2$ under pressure have been investigated by Wang *et al.*[35] and various other physical properties of $Ti_2InB_2$ have been studied by Ali *et al.*[36]. In the meantime, Miao *et al.* [30] have examined possible 314 and 212 MAX phase borides and checked their thermodynamic stability. The 314 MAX phase $Hf_3PB_4$ has further been investigated by DFT and predicted as the hardest possible MAX phase compound known so far [27]. The physical properties of 212 MAX phase borides, $Hf_2AB_2$ (A = In, Sn), have been disclosed [28] and found to have improved thermo-mechanical properties compared to other relevant MAX phase boride/carbides. The structure of 212 phases is slightly different from 211 MAX phases. In this case (212 phases), there is a 2D layer of B sandwiched between M layers in which an extra B atom exists at the position of X unlike in 211 phases. Within the 2D layer, two center-two electrons (2c-2e) bond is formed that leads strong covalent B-B bond [28–30,35,37]. This B-B bonding results in improved mechanical and thermal properties of 212 MAX phase borides. It is also observed that the mechanical properties of 211 MAX phase borides are slightly different (degraded/upgraded) from the corresponding carbides as reported by different groups [11,22,38–41]. The aforementioned reports inspired us to select the predicted MAX phase borides $Zr_2AB_2$ (A = In, Tl) for further study. The successful synthesis of $Ti_2InB_2$ made us optimistic about the successful synthesis of $Zr_2AB_2$ (A = In, Tl) in future and motivated us for detailed calculations of the physical properties of the titled MAX phase borides. The structural aspects of these two MAX phase borides are similar to those of $Ti_2InB_2$ and $Hf_2AB_2$ (A = In, Sn) MAX phases and they also belong to the class of 212 MAX phase borides.

Miao *et al.*[30] have theoretically confirmed the stability of $Zr_2AB_2$ (A = In, Tl) from structural, thermodynamic, and chemical points of view. Information regarding the physical properties of these compounds is important to judge their potential for future applications. Since MAX phases are potential materials for practical applications in many sectors; exploration of the physical properties of $Zr_2AB_2$ (A = In, Tl) carries significant scientific and technological importance. Keeping this in mind, we have carried out the present study.

The hitherto unexplored physical properties such as mechanical properties, elastic anisotropy, details of electronic structure, important thermal and optical properties of $Zr_2AB_2$ (A = In, Tl)

MAX phase borides have been presented and discussed in this paper. The studied properties are compared with those of other 212 MAX phase borides, where available.

## 2. Computational methodology

The calculations of the aforementioned properties of $Zr_2AB_2$ (A = In, Tl) are performed using density functional theory based on the plane-wave pseudopotential method as implemented in the CAmbridge Serial Total Energy Package (CASTEP) code [42,43]. The exchange and correlation functions were treated by the generalized gradient approximation (GGA) of the Perdew–Burke–Ernzerhof (PBE) [44]. The pseudo-atomic calculations were performed for B - $2s^2\ 2p^1$, In- $5s^2 5p^1$, Tl-$6s^2\ 6p^1$ and Zr – $5s^2\ 4d^2$ electronic orbitals. For integration within theBrillouin zone, a k-point [45]mesh of size 10 × 10 × 4 was selected and the cutoff energy wasset to 500 eV. The relaxation of the structures was done by Broyden Fletcher Goldfarb Shanno (BFGS) technique [46] and the electronic structure was calculated using density mixing. The parameters for relaxed structures have the following tolerance thresholds: the self-consistent convergence of the total energy is $5 \times 10^{-6}$ eV/atom, the maximum force on the atom is 0.01 eV/Å, the maximum ionic displacement is set to $5 \times 10^{-4}$ Å with a maximum stress of 0.02 GPa.

## 3. Results and discussion

### 3.1 Structural properties and stability

Fig. 1 shows the unit cell structure of $Zr_2InB_2$ (212 MAX phase) as a representative of $Zr_2AB_2$ (A = In, Tl) MAX phases. The structure of $Zr_2AB_2$ (A = In, Tl) is slightly different from the conventional C and N containing MAX phases. The space group of $Zr_2AB_2$ (A = In, Tl) is $P\bar{6}m2$, (No. 187) [30] whereas the space group of conventional of MAX phase is $P63/mmc$ (No. 194) [13]. The unit cell of a 211 MAX phase ($Zr_2InC$) is also presented along with that of 212 MAX phases so that the difference between the structures can easily be understood. For 211 MAX phases, the atomic positions of M (Zr), A (In) and X (C) atoms in the unit cell are (1/3, 2/3, $z_M$), (1/3, 2/3, 3/4) and (0, 0, 0). On the other hand, there are two X atoms (B) which are positioned at (0.6667, 0.3333, 0.5) and (0.0, 0.0, 0.5) in the boride compound. The M (Zr) atom is located at (0.3333, 0.6667, 0.6935)and A (In/Tl) is positioned at (0.6667, 0.3333, 0.0). The X elements are positioned at the corners of the unit cell for 211 MAX phases while the X elements form a 2D layer between the M layers in 212 MAX phases. Covalent B-B bonding is formed within the 2D layer that leads to a more stable structure compared to conventional 211 MAX phases. The cell parameters of relaxed structures are presented in Table 1. The obtained values of the lattice parameters are very much consistent with the prior results [30] that validate the correctness of the computational methodology for present calculations. For example, very low deviations: 0.08% (0.03%), 0.09% (0.016%) and 0.27% (0.05%) for $a$, $c$, and $V$, respectively of $Zr_2InB_2$ ($Zr_2TlB_2$) from previous measures [30] are noticed.

**Table 1:** Calculated lattice parameters ($a$ and $c$), $c/a$ ratio, and volume (V) of $Zr_2AB_2$ (A = In, Tl) MAX phases.

| Phase | $a$ (Å) | % of deviation | $c$ (Å) | % of deviation | $c/a$ | $V$ (Å$^3$) | % of deviation | Reference |
|---|---|---|---|---|---|---|---|---|
| $Zr_2InB_2$ | 3.2358 | 0.08 | 8.4933 | 0.09 | 2.608 | 77.0160 | 0.27 | This study |
|  | 3.233 |  | 8.485 |  | 2.611 | 76.8035 |  | Theo. [30] |
| $Zr_2TlB_2$ | 3.2549 | 0.03 | 8.4855 | 0.016 | 2.543 | 77.8564 | 0.05 | This study |
|  | 3.254 |  | 8.486 |  | 2.547 | 77.8137 |  | Theo. [30] |

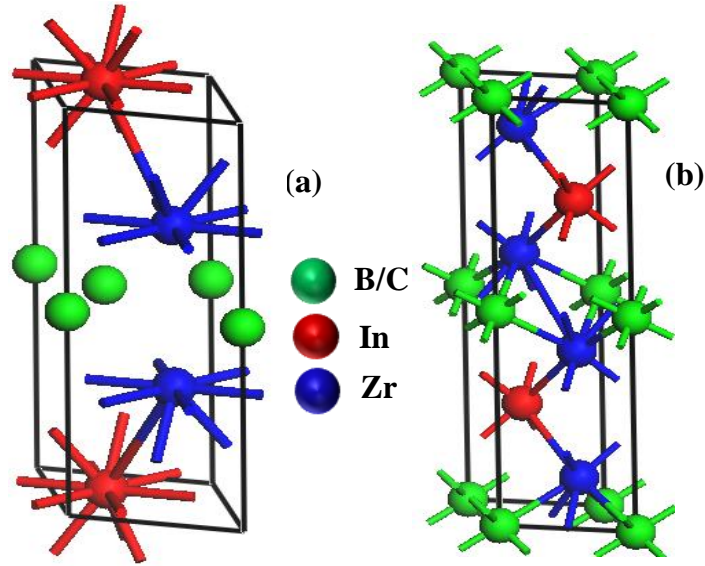

Fig. 1: The schematic unit cell of (a) $Zr_2InB_2$ and (b) $Zr_2InC$ MAX phase nanolaminates.

It is scientifically significant that the structure obtained either experimentally or theoretically should be stable, thus, checking the stability of a crystal is routine work. We have checked the stability using two approaches: (i) dynamical stability by calculating the phonon dispersion curves and (ii) mechanical stability by calculating the stiffness constants ($C_{ij}$).

To judge the dynamic stability of the MAX phase borides under consideration, phonon dispersion curves (PDCs) at ground state have been calculated using the Density Functional Perturbation Theory (DFPT) linear-response method [47]. The PDC along the high symmetry directions of the crystal Brillouin zone (BZ) together with the total phonon density of states (PHDOS) of $Zr_2AB_2$ (A = In, Tl)compounds are presented in Figs. 2 (a, c). Existence of only positive frequencies in the PDC suggests the dynamical stability whereas the negative frequency branch at any *k*-point indicates the dynamical instability of a compound. No negative frequency branch has been observed in the displayed PDCs in Figs. 2 (a, c), consequently the compounds $Zr_2AB_2$ (A = In, Tl) are predicted to be dynamically stable.

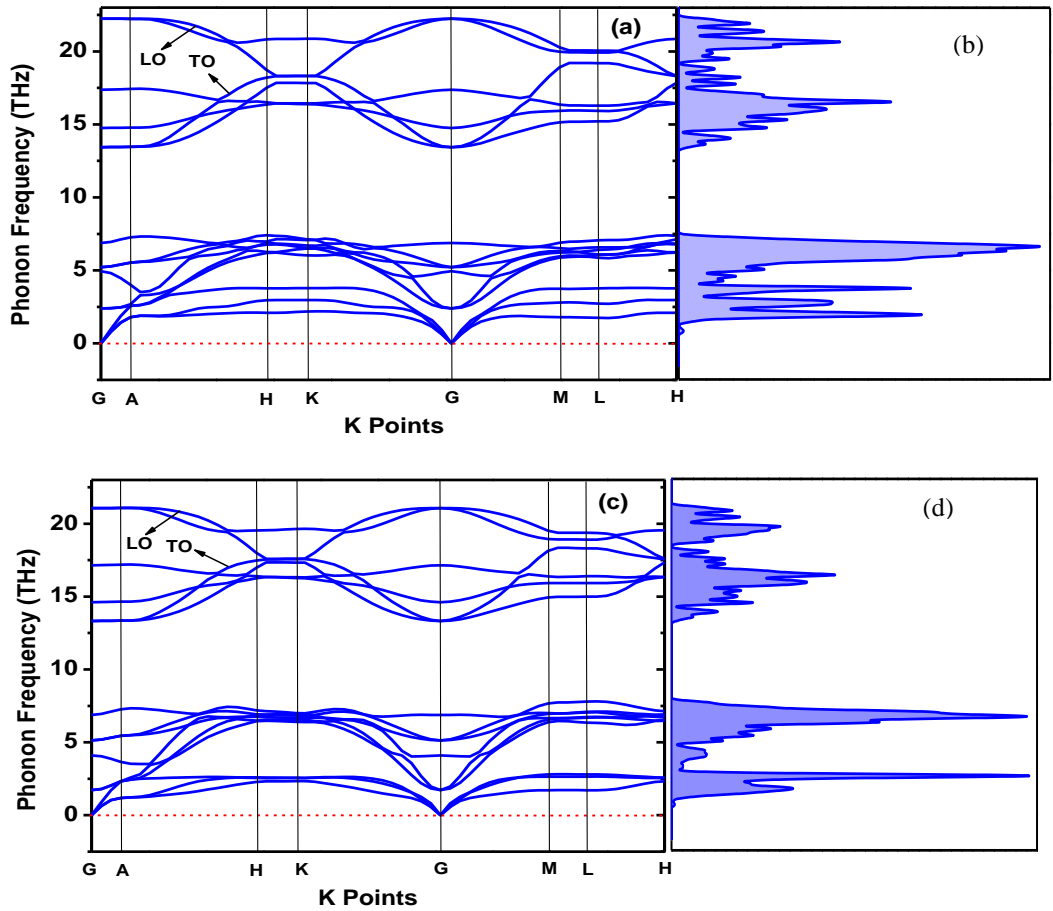

Fig. 2: The phonon dispersion curves (PDCs) and phonon density of states (PHDOS) of $Zr_2InB_2$ (a, b), and $Zr_2TlB_2$ (c, d) compounds.

The PHDOS of the compounds $Zr_2AB_2$ (A = In, Tl) are calculated from the PDCs and displayed side by side along with the PDCs for better understanding in Figs. 2(b, d) and to identify bands by comparing the corresponding peaks. The flatness of the bands for TO shown in Fig. 2 (a) gives a pronounced peak [Fig. 2(b)] while non-flat bands for LO in Fig. 2(a) results in weak peaks [Fig. 2(b)] in the PHDOS. Similar links are found between Fig. 2(c) and Fig. 2(d). It is noteworthy that a clear gap exists between the acoustic and optical branches. In addition, the top of the LO and bottom of the TO modes are located at the Γ point and the separation between these is found to be 5.8 and 7.87 THz for $Zr_2InB_2$ and $Zr_2TlB_2$ compounds, respectively. The confirmation of mechanical stability using stiffness constants has been presented in the next section.

### 3.2 Mechanical properties

*3.2.1 Stiffness constants and elastic moduli*

During practical use, materials are subjected to forces or loads of different nature. Therefore, it is essential to have knowledge about the characteristics of the material to avoid any kind of permanent deformation or fracture. The mechanical properties of solids reveal itself in the reaction to deformation under the application of a load or force. Key parameters that characterize the mechanical properties of materials are stiffness constants, strength, elastic moduli, hardness, ductility/brittleness, machinability, and toughness. In this section we investigate about the mentioned properties to disclose the mechanical behavior of $Zr_2AB_2$ (A = In, Tl) MAX phase borides. The stiffness constants that are basis to the study of mechanical properties have been calculated through the strain-stress method as contained within the CASTEP program [35,48–51]. The estimated stiffness constants of $Zr_2AB_2$ (A = In, Tl) are shown in Table 2. No comparison with prior results is possible at this time. In this situation, comparison with those of corresponding carbides is of scientific interest. The stiffness constants of 211 MAX phase carbides are therefore presented in Table 2 for comparison. The stiffness constants are used to check the mechanical stability crystalline solids. The conditions proposed by Max Born [52] on the stiffness constants certify the mechanical stability which are mostly applicable for cubic crystalline solids and are not sufficient for other crystal systems. Mouhat *et al.*[53] reviewed the stability conditions for all crystal symmetries. Accordingly, the conditions for mechanical stability of a hexagonal system becomes: $C_{11} > 0$, $C_{11} > C_{12}$, $C_{44} > 0$, $(C_{11} + C_{12})C_{33} - 2(C_{13})^2 > 0$. As listed in Table 2, the $C_{ij}$ values of $Zr_2AB_2$ (A = In, Tl) satisfy the above conditions and theoretically confirmed the mechanical stability of $Zr_2AB_2$ (A = In, Tl). The $C_{ij}$ are also used to understand the origin of elastic anisotropy within these borides. Recalling Fig. 1, the atomic arrangements are different along different crystallographic axes. Due to the differences in the atomic arrangements, the bonding strengths are also different for the same which are reflected from the unequal values of $C_{11}$ and $C_{33}$. It is well known that for hexagonal system $C_{11}$ measures the required resistance for deformation along the $a(b)$-axis whereas $C_{33}$ indicates the same along the $c$-axis. The higher value of $C_{11}$ than that of $C_{33}$ for $Zr_2AB_2$ (A = In, Tl) borides reveal the higher bonding strength, stiffness and resistance to deformation along the $a(b)$-axis compared to those along the $c$-axis which is mainly responsible for the elastic anisotropy possessed by these MAX phase borides.

**Table 2:** The elastic stiffness constants, $C_{ij}$ (GPa), bulk modulus, $B$ (GPa), shear modulus, $G$ (GPa), Young's modulus, $Y$ (GPa), $H_{macro}$, $H_{micro}$ (GPa), Pugh ratio, $G/B$, Poisson ratio, $v$ and Cauchy Pressure, $CP$ (GPa) of $Zr_2AB_2$ (A = In, Tl), together with those of $Zr_2AC$ (A = In, Tl), $Hf_2AB_2$ (A = In, Sn), $Hf_2AC$ (A = In, Sn), $Ti_2InB_2$ and $Ti_2InC$ MAX compounds.

| Phase | $C_{11}$ | $C_{12}$ | $C_{13}$ | $C_{33}$ | $C_{44}$ | $B$ | $G$ | $Y$ | $G/B$ | $H_{macro}$ | $H_{micro}$ | $v$ | Cauchy Pressure | Reference |
|---|---|---|---|---|---|---|---|---|---|---|---|---|---|---|
| $Zr_2InB_2$ | 315 | 46 | 64 | 263 | 82 | 138 | 105 | 251 | 0.76 | 19.11 | 21.24 | 0.20 | -36 | This |
| $Zr_2InC$ | 286 | 62 | 71 | 248 | 83 | 136 | 95 | 231 | 0.70 | 15.87 | 17.94 | 0.22 | -14 | Ref-[54] |
| $Zr_2TlB_2$ | 310 | 52 | 61 | 251 | 66 | 135 | 94 | 229 | 0.70 | 15.68 | 17.71 | 0.22 | -21 | This |
| $Zr_2TlC$ | 255 | 60 | 52 | 207 | 63 | 115 | 80 | 195 | 0.70 | 13.98 | 15.06 | 0.22 | -03 | Ref-[55] |
| $Hf_2InB_2$ | 343 | 61 | 76 | 278 | 94 | 154 | 114 | 274 | 0.74 | 19.46 | 22.56 | 0.20 | -33 | Ref-[28] |
| $Hf_2InC$ | 309 | 81 | 80 | 273 | 98 | 152 | 105 | 256 | 0.69 | 16.75 | 19.65 | 0.21 | -17 | Ref-[54] |
| $Hf_2SnB_2$ | 353 | 65 | 86 | 306 | 110 | 165 | 124 | 297 | 0.75 | 21.02 | 24.84 | 0.20 | -45 | Ref-[28] |
| $Hf_2SnC$ | 251 | 71 | 107 | 238 | 101 | 145 | 87 | 218 | 0.60 | 12.00 | 14.50 | 0.25 | -30 | Ref-[56] |
| $Ti_2InB_2$ | 364 | 47 | 58 | 275 | 94 | 147 | 122 | 287 | 0.83 | 23.72 | 26.44 | 0.17 | -47 | Ref-[28] |
| $Ti_2InC$ | 284 | 62 | 51 | 242 | 87 | 126 | 100 | 236 | 0.79 | 19.57 | 20.92 | 0.18 | -25 | Ref-[36] |

Table 2 also shows the bulk modulus ($B$) and shear modulus ($G$) that are calculated using Hill's approximation [57,58]. Hill's values are the average values of the upper limit (Voigt [59]) and lower limit (Reuss[60]) of $B$: $[B = (B_V + B_R)/2]$ and $G$: $[G = (G_V + G_R)/2]$. The $B_V$, $B_R$, $G_V$ and $G_R$ have been calculated from the stiffness constants by the following relations: $B_V = [2(C_{11} + C_{12}) + C_{33} + 4C_{13}]/9$; $B_R = C^2/M$; $C^2 = C_{11} + C_{12})C_{33} - 2C_{13}^2$; $M = C_{11} + C_{12} + 2C_{33} - 4C_{13}$; $G_V = [M + 12C_{44} + 12C_{66}]/30$ and $G_R = \left(\frac{5}{2}\right)[C^2 C_{44} C_{66}]/[3B_V C_{44} C_{66} + C^2(C_{44} + C_{66})]$; $C_{66} = (C_{11} - C_{12})/2$. The $B$ and $G$ of $Zr_2AB_2$(A = In, Tl) are larger than those of corresponding 211 MAX phase carbides $Zr_2AC$ (A = In, Tl). Similar results are also found for other 212 MAX phase borides and corresponding 211 carbides as presented in Table 2. Thus, it can be concluded that the resistance to volume and plastic deformations are higher for 212 MAX phase borides than 211 carbides. The elastic modulus that provides stiffness of solids is the Young's modulus ($Y$) and can be calculated using the relation: $Y = 9BG/(3B + G)$[51,61]. The obtained values of $Y$ also confirm that the 212 MAX phases are stiffer than their 211 MAX phase counterparts. Altogether, these three elastic moduli ($B$, $G$ and $Y$) do not measure hardness directly but provide some useful prior information regarding hardness. It is easy to predict the hardness variation among a group of solids from their elastic moduli. Thus, it is reasonable to expect that the hardness of $Zr_2AB_2$ (A = In, Tl) would be greater than that of $Zr_2AC$ (A = In, Tl). Similar results are already reported for $Hf_2AB_2$ (A = In, Sn) and $Hf_2AC$ (A = In, Sn); $Ti_2InB_2$ and $Ti_2InC$[28]. Among these three elastic moduli, $G$ is assumed to be closely related to the hardness of solids. It is noticed that the value $G$ is increased by 10.5% and 17.5% for $Zr_2AB_2$(A = In, Sn), respectively, compared to $Hf_2AC$ (A = In, Sn). It has been also shown that $G$ is found to be higher by 8.5%, 42.5% and 22% for $Hf_2AB_2$ (A = In, Sn) and $Ti_2InB_2$, respectively compared to $Hf_2AC$ (A = In, Sn) and $Ti_2InC$. In case of 314 MAX phase ($Hf_3PB_4$), $G$ is noticed to be increased by 41.7%, 41.7% and 63.6% compared to other Hf based 312 MAX phases $Hf_3AC_2$ (A = Al, S, Sn), respectively. The predicted hardness of $Hf_3PB_4$ is the highest among all the known MAX compounds so far. To elucidate the matter further and to understand the effect of double X atoms on the hardness, we have calculated different hardness parameters of $Zr_2AC$ (A = In, Tl) in the next section.

### 3.2.2 Hardness of $Zr_2AB_2$ (A = In, Tl)

This section is dedicated to the hardness parameters of $Zr_2AC$ (A = In, Tl). Hardness of a solid can be understood from different theoretical schemes. Like experimental techniques employed to measure hardness, these different formalisms may also give different values of theoretical hardness. Two hardness parameters known as macro- and micro-hardness have been calculated using the elastic moduli from the following relations: $H_{micro} = \frac{(1-2\nu)E}{6(1+\nu)}$[62] and $H_{macro} = 2\left[\left(\frac{G}{B}\right)^2 G\right]^{0.585} - 3$ [63]. As seen in Table 2, the $H_{macro}$ and $H_{micro}$ are higher for 212 MAX phase borides than those of 211 MAX phase carbides. For $Zr_2InB_2$, the $H_{macro}$ and $H_{micro}$ are 20.4% and 20.0% larger than those of $Zr_2InC$. While for $Zr_2TlB_2$, the $H_{macro}$ and $H_{micro}$ are 12.1% and 17.6% larger than those of $Zr_2TlC$. The $H_{macro}$ ($H_{micro}$) is 16.2% (14.8%), 75.2% (71.3%) and 21.2% (26.4%) larger for $Hf_2InB_2$, $Hf_2SnB_2$ and $Ti_2InB_2$ than those of $Hf_2InC$, $Hf_2SnC$ and $Ti_2InC$.

Furthermore, the $H_{macro}$ ($H_{micro}$) of $Hf_3PB_4$ is 29.0% (44.0%), 59.8% (63.7%) and 84.43% (94.1%) higher than those of $Hf_3AC_2$ (A = Al, S, Sn), respectively. In fact, the $H_{macro}$ and $H_{micro}$ of $Hf_3PB_4$ are higher than those of all known MAX phase nanolaminates. Some 211 MAX phase borides: $M_2SB$ (M = Zr, Hf, Nb) has been already synthesized. The hardness $H_{macro}$ and $H_{micro}$ of $M_2SB$ [M = Zr (16.37 and 18.67 GPa), Hf (18.17 and 2.63 GPa)] are lower than those of $M_2SC$ [M = Zr (17.79, 21.57 GPa), Hf (17.24, 23.88 GPa)] while $H_{macro}$ and $H_{micro}$ of $Nb_2SB$ (18.76 and 19.10 GPa) is higher than those of $Nb_2SC$ (11.58 and 15.84 GPa); implying that the hardness of MAX compounds will be either increased or decreased when B simply replaces C. Whereas, the replacement of C by B and doubling of X atoms in the 212 MAX phases leads to an increase in hardness more or less. Gou *et al.* [64] have proposed a formula for partially metallic bonded compounds that is suitable to calculate the Vickers hardness ($H_v$) based on the Mulliken bond populations. The relevant formula for the hardness is given as [65]:

$$H_V = \left[\prod^{\mu}\left\{740(P^{\mu} - P^{\mu'})(v_b^{\mu})^{-5/3}\right\}^{n^{\mu}}\right]^{1/\sum n^{\mu}}$$ here $P^{\mu}$ is the Mulliken population of the $\mu$-type bond,

$P^{\mu'} = n_{free}/V$ is the metallic population, and $v_b^{\mu}$ is the bond volume of $\mu$-type bond. The obtained values of $H_v$ for both titled compounds are given Table 3. The $H_v$ of $Zr_2InB_2$ (2.92 GPa) is 59.6% higher than that of $Zr_2InC$ (1.053 GPa) [7]. No report is available for $Zr_2TlC$, thus, comparison is not possible at this stage. Since, the elastic moduli of $Zr_2TlC$ are smaller than those of $Zr_2InC$; consequently, a lower value of $H_v$ for $Zr_2TlC$ is expected and a significant increase in $H_v$ is also expected for $Zr_2TlB_2$. The $H_v$ of $Hf_2AB_2$ [A = In (3.94 GPa), Sn (4.41 GPa)] and $Ti_2InB_2$ (4.05 GPa) are 14.2%, 16.1% and 55.7% higher than that of $Hf_2AC$ [A = In (3.45 GPa), Sn (3.80 GPa)] and $Ti_2InC$ (2.60 GPa). Moreover, The $H_v$ of $Hf_3PB_4$ is 7.85 GPa which is much higher than that of $Hf_3AlC_2$ (4.9 GPa) [66] and $Hf_3SnC_2$ (4.7 GPa) [67].

Now, a reasonable question can arise that why are the mechanical properties of 212 MAX phase borides significantly higher than those of 211 MAX phase carbides? We believe, the answer is present within the structure of 212 MAX phase boride compounds. As shown in Fig. 1 (a), there is a 2D layer of B atoms sandwiched between the M atoms. The B atoms form a very strong B-B bond by sharing two center-two electrons (2c-2e). One of the simple approaches for understanding the bonding strength is to look at the bond overlap population (BOP) of Mulliken population analysis. The bonding strength is usually proportional to the BOP. For 211 MAX phases, the highest BOP is found for M-X bond which is usually low, for example, one gets 1.07, 1.41 and 1.04 for $Zr_2InC$, $Hf_2InC$ and $Ti_2InC$, respectively. Whereas for 212 MAX phase borides the highest BOP is found for B-B bond formed within the 2D boron layer. The BOP is usually higher for this bond, such as 2.22 ($Zr_2InB_2$), 2.17 ($Zr_2TlB_2$), 2.38 ($Hf_2InB_2$), 2.26 ($Hf_2SnB_2$), 2.34 ($Ti_2InB_2$). Although the BOP for M-X bond is higher for 211 MAX phases but for 212 MAX phases it is very low compared to the B-B bond. The geometrical averages of the bonding strength remain higher for 212 MAX phases compared to the 211 MAX phases that lead to higher Vickers hardness for 212 MAX phases. The evaluation bonding strength can be visualized in terms of charge density mapping. Fig. 3 shows the charge density mapping of (a) $Zr_2InB_2$ and (b) $Zr_2TlB_2$ in which the red color indicate the highest electron density and the blue color indicates the lowest electron density. As seen in Fig. 3, the highest electron density is observed at B atomic position. Mulliken analysis confirmed that 0.57 $|e|$ charge is transferred from Zr atom to B atoms (located both at edge and interstitial positions, [Fig. 1 (a)]). The charges are then further shared among the B atoms within 2D layers and resulted in a very strong covalent bond by making a 2c-2e bond. The strength of this bond is much higher than other bonds such as B-Zr bonds which are reflected from the values BOP shown in Table 3. In the case of $Zr_2InB_2$, no significant bond is formed between Zr and In atoms due to the long bond length between them. On the other hand, the charges are accumulated at C atomic positions that come from the Zr atoms and form the strongest covalent bonds compared to other bonds present within the $Zr_2InC$ compound [7]. The BOP is 1.07 and 0.11 for Zr-C and Zr-In bonds within $Zr_2InC$. Now, based on the analysis regarding charge density mapping and BOP it is clear that the mechanical properties of 212 MAX phases are significantly enhanced due to the presence of a 2D layer of B atoms within their structure.

**Table 3:** Calculated Mulliken bond number $n^\mu$, bond length $d^\mu$, bond overlap population $P^\mu$, metallic population $P^{\mu'}$, bond volume $v_b^\mu$, bond hardness $H_v^\mu$ of $\mu$-type bond and Vickers hardness $H_v$ of $Zr_2AB_2$ (A = In, Tl).

| Compounds | Bond | $n^\mu$ | $d^\mu$ (Å) | $P^\mu$ | $P^{\mu'}$ | $v_b^\mu$ (Å$^3$) | $H_v^\mu$ (GPa) | $H_v$ (GPa) |
|---|---|---|---|---|---|---|---|---|
| $Zr_2InB_2$ | B1-B2 | 1 | 1.86821 | 2.22 | 0.0452 | 07.187 | 60.12 | 2.92 |
| | B1-Zr1 | 2 | 2.51124 | 0.21 | 0.0452 | 17.457 | 1.038 | |
| | B2-Zr2 | 2 | 2.51124 | 0.20 | 0.0452 | 12.012 | 1.818 | |
| $Zr_2TlB_2$ | B1-B2 | 1 | 1.87924 | 2.17 | 0.04278 | 07.373 | 56.35 | 2.19 |
| | B1-Zr1 | 4 | 2.51247 | 0.20 | 0.04278 | 17.621 | 0.975 | |

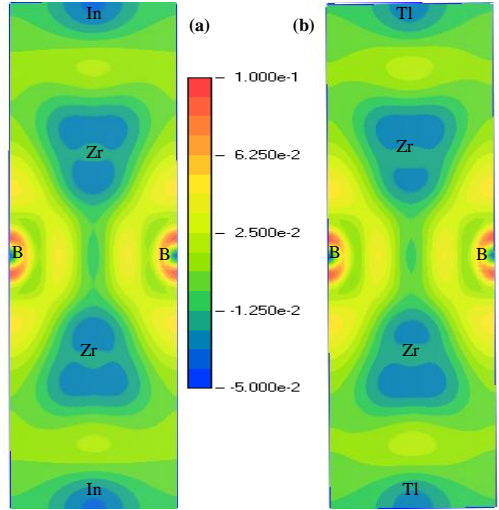

Fig. 3: The charge density mapping of $Zr_2AB_2$ (A = In, Tl) MAX phase borides.

### *3.2.3 The brittleness of $Zr_2AB_2$ (A = In, Tl)*

In this section, we used three models to predict whether $Zr_2AB_2$ (A = In, Tl) borides will be ductile or brittle. The ratio of shear modulus to bulk modulus (Pugh ratio, *G/B*) [85] corresponds to the contest between two processes, plasticity and fracture. For ductile solids plasticity is easier while the fracture is easier for brittle solids. The transition from ductile to brittle is judged by the critical value of *G/B* = 0.571. A low value of the *G/B* ratio implies the first (ductile) type of solids, while a high value of the *G/B* ratio implies the second type (brittle) of solids. As evident from the values given in Table 2, the $Zr_2AB_2$ (A = In, Tl) compounds are brittle materials. Value of Poisson's ratio (*υ*) can also separate the brittle and ductile materials with a knowledge of the critical value of 0.26 [66]. The value $υ < 0.26$ implies brittleness and $υ > 0.26$ implies ductility. The Poisson's ratio is computed using the equation: $υ = (3B − Y)/(6B)$ [51,61]. Like the Pugh ratio, the Poisson's ratio also certifies that the $Zr_2AB_2$ (A = In, Tl) compounds are brittle materials. Moreover, value of *υ* provides information regarding bonding nature within the materials where a small value of *υ* (~ 0.10) indicates the covalent bonding and a high value of *υ* (~ 0.33) indicates the metallic bonding. The values of *υ* for $Zr_2AB_2$ (A = In, Tl) lie between the two extremes, indicating a mixture covalent/ionic/metallic bondings within these solids. Pettifor [68] discussed the Cauchy pressure (*CP*) in determining the bonding as well as the failure mode of solids. The *CP* is defined in terms of stiffness constants where $CP = (C_{12} - C_{44})$. If a material has a high resistance tobond bending which is also covalently bonded then the *CP* of that materials will be negative i.e., $C_{44}$ will be higher than that of $C_{12}$. In contrast for metallic bonded solid, the *CP* will be positive i.e., $C_{12} > C_{44}$. Pettifor [68] also suggests that *CP* will appeal to the Pugh ratio when commenting on the failure mode of solids. *CP* with a positive value implies ductile solids and reverse is true for brittle solids. The values of *CP* for $Zr_2AB_2$ (A = In, Tl) again categorize them as brittle materials. Finally, all the results regarding the analysis of failure mode suggest that the studied borides are brittle in nature like most other MAX phase carbides and nitrides.

### *3.2.4 Mechanical anisotropy of $Zr_2AB_2$ (A = In, Tl)*

It is evident from a large number of previous investigations that the MAX phases are prospective materials for technological applications. Thus, detailed knowledge of mechanical anisotropy is a prerequisite for many technological sectors because of its close relation with physical processes like crack formation and propagation, anisotropic plastic deformation, and unusual phonon modes, etc. [69,70]. The mechanical anisotropy can be judged by calculating the different anisotropy indices from the stiffness constants and via graphical presentations of the elastic moduli in 2D and 3D.

The graphical presentations of Young's modulus, compressibility, shear modulus, and Poisson's ratio are carried out via the ELATE code [71] as demonstrated in Fig. 4 (a-d) for $Zr_2InB_2$ and Fig. 5 (a-d) for $Zr_2TlB_2$. The 2D and 3D plots of elastic moduli are very effective as a perfect circle (for 2D) and sphere (for 3D) represent isotropy while deviations from these shapes represent the anisotropy where the degree of deviation corresponds to the level of anisotropy.

The anisotropy of the elastic moduli for $Zr_2AB_2$ (A = In, Tl) is similar in nature except for the extent of their anisotropy. For example, the value of *Y* is maximum on the horizontal axis for both phases while it is minimum on the vertical axis in *xz* and *yz* planes. The *Y* is isotopic in *xy* plane for both phases. The $Zr_2TlB_2$ is more anisotropic compared to $Zr_2InB_2$ as evident from the values [Table 4] of $A_Y$ and the figures [4 (a) and 5 (a)] as well. A reverse nature is observed for the anisotropy in compressibility (*K*) in *xz* and *yz* planes where the largest value of *K* is noted on the vertical axis and the smallest value is noted on the horizontal axis. The degree of anisotropy in *K* is also greater in $Zr_2TlB_2$ compared to $Zr_2InB_2$ [Table 4; Fig. 4 (b) and 5 (b)] and *K* is isotropic in the *xy* plane. A different anisotropic nature is noted for the shear modulus (G) and Poisson's ratio (υ) for both phases as shown in Fig. 4 (c and d) and 5 (c and d). The maximum of *G* is represented by blue color and the minimum is represented by green color at any points. The $G$ is maximum on the horizontal axes as presented by the blue line and minimum on the vertical axis where both blue and green lines coincide in $xz$ and $yz$ planes. The anisotropy of $v$ is demonstrated in Fig. 4 (d) and 5 (d) where the $v$ is observed to be maximum (blue line) at an angle of 45° between both horizontal and vertical axes and minimum on the horizontal axes as presented by the green line. The minimum and the maximum values of the elastic moduli along different directions and their ratio ($A$) are tabulated in Table 4. Table 4 also contains the same for the $Hf_2AB_2$ (A = In, Sn) and $Ti_2InB_2$ compounds for comparison.

**Table 4:** The minimum and the maximum values of the Young's modulus, compressibility, shear modulus, and Poisson's ratio of $Hf_2AB_2$ (A = In, Sn), together with those of the $Ti_2InB_2$ for comparison.

| Phase | $Y_{min.}$ (GPa) | $Y_{max.}$ (GPa) | $A_Y$ | $K$ ($TPa^{-1}$) | $K$ ($TPa^{-1}$) | $A_K$ | $G_{min.}$ (GPa) | $G_{max.}$ (GPa) | $A_G$ | $v_{min.}$ | $v_{max.}$ | $A_v$ |
|---|---|---|---|---|---|---|---|---|---|---|---|---|
| $Zr_2InB_2$ | 218.17 | 296.17 | 1.536 | 2.2921 | 2.6738 | 1.166 | 82.76 | 134.52 | 1.625 | 0.100 | 0.334 | 3.314 |
| $Zr_2TlB_2$ | 184.43 | 290.41 | 1.575 | 2.2773 | 2.8713 | 1.261 | 65.63 | 129.17 | 1.968 | 0.104 | 0.415 | 3.970 |
| $Hf_2InB_2$ | 240.23 | 316.92 | 1.32 | 2.000 | 2.500 | 1.25 | 94.223 | 140.81 | 1.49 | 0.125 | 0.308 | 2.46 |
| $Hf_2SnB_2$ | 270.57 | 323.40 | 1.19 | 1.948 | 2.175 | 1.11 | 110.46 | 143.66 | 1.30 | 0.125 | 0.267 | 2.13 |
| $Ti_2InB_2$ | 240.80 | 249.11 | 1.45 | 2.042 | 2.775 | 1.36 | 094.13 | 158.89 | 1.69 | 0.095 | 0.309 | 3.25 |

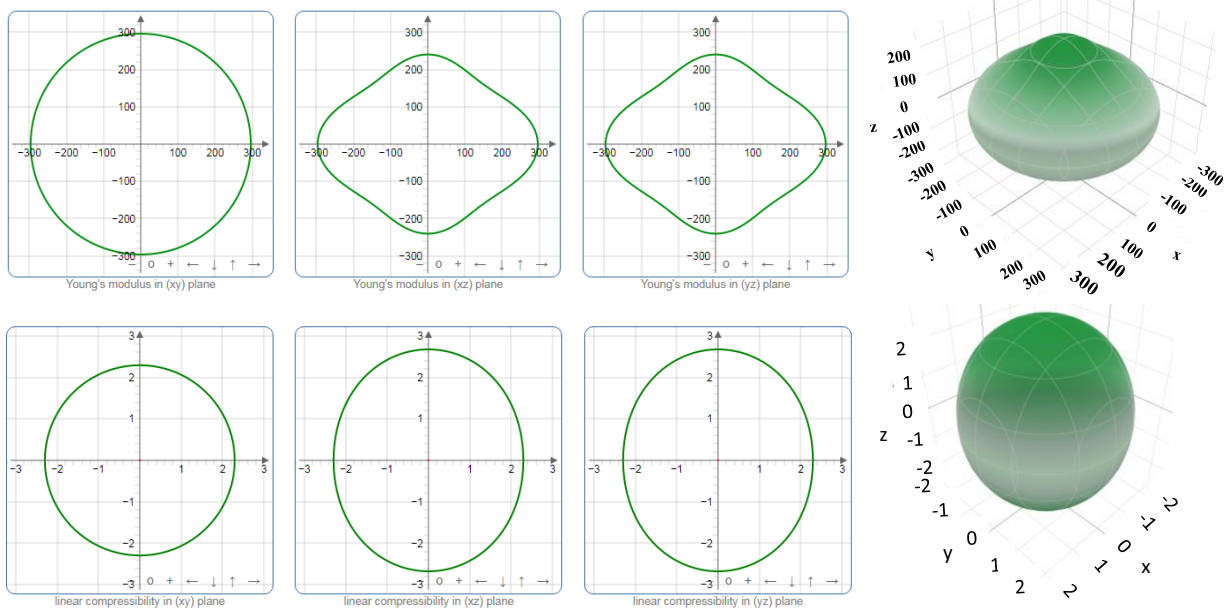

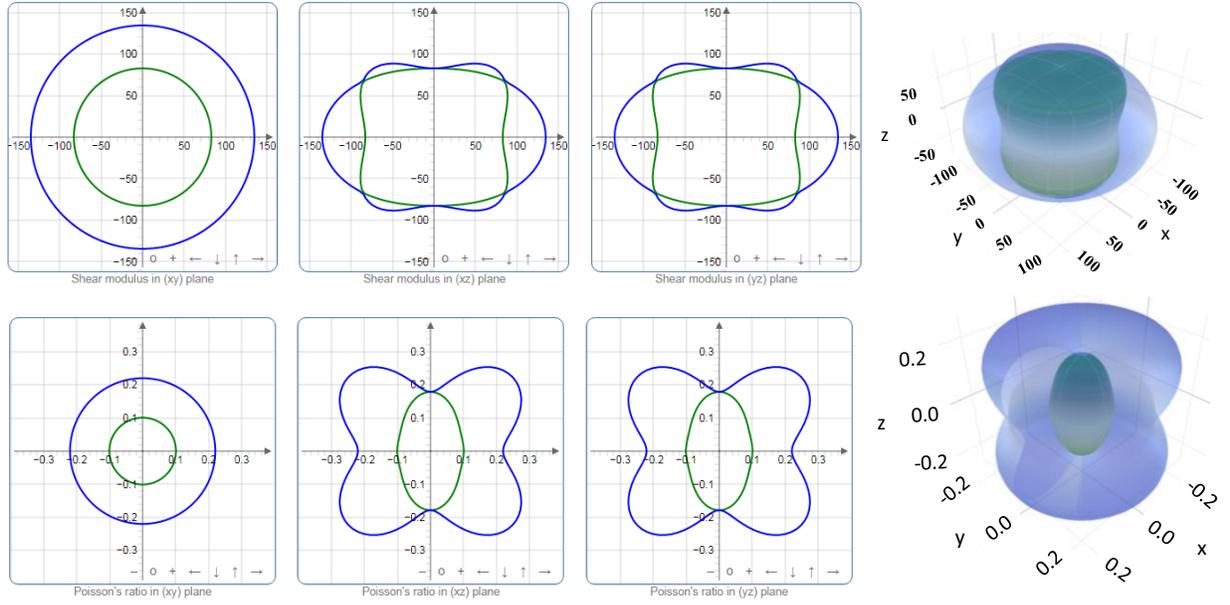

Fig. 4: The 2D and 3D plots of (a) *Y*, (b) *K*, (c) *G* and (d) *υ* of Zr$_2$InB$_2$.

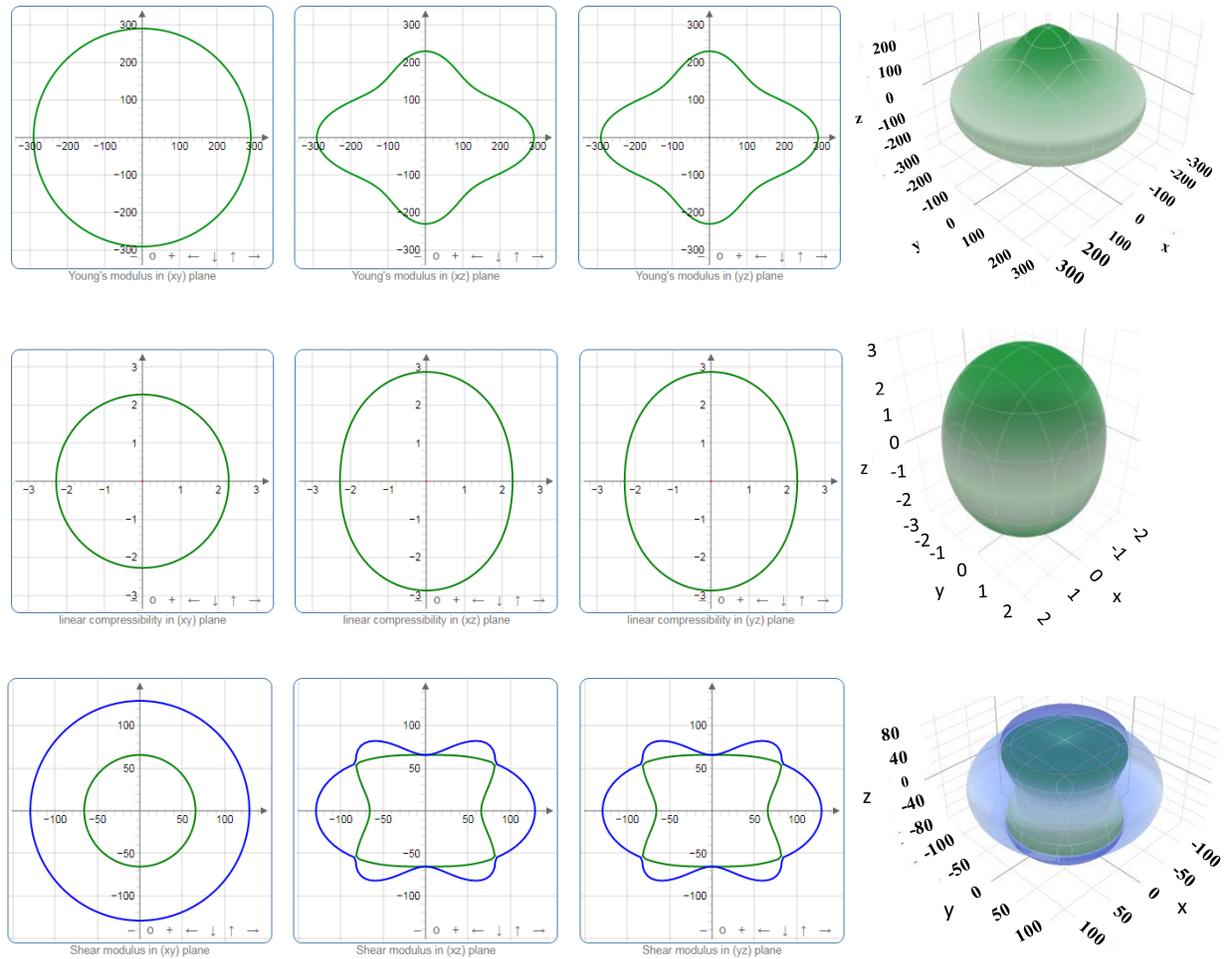

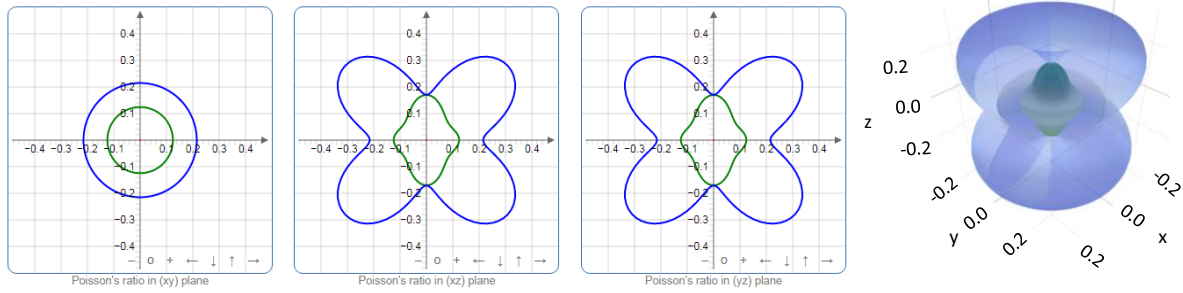

Fig. 5: The 2D and 3D plots of (a) $Y$, (b) $K$, (c) $G$ and (d) $v$ of $Zr_2TlB_2$.

The different anisotropy indices have also been calculated using the formulae which can be found elsewhere [22] and are presented in Table 5.

**Table 5:** Anisotropy factors, $A_1$, $A_2$, $A_3$, $k_c/k_a$, $B_a$, $B_c$ and universal anisotropy index $A^U$ of $Zr_2AB_2$ (A = In, Tl), together with those of $Hf_2AB_2$ (A = In, Sn) and $Ti_2InB_2$ for comparison.

| Phase | $A_1$ | $A_2$ | $A_3$ | $k_c/k_a$ | $B_a$ | $B_c$ | $A^U$ |
|---|---|---|---|---|---|---|---|
| $Zr_2InB_2$ | 1.28 | 0.61 | 0.78 | 1.17 | 382 | 537 | 0.246 |
| $Zr_2TlB_2$ | 1.57 | 0.51 | 0.80 | 1.26 | 383 | 498 | 0.496 |
| $Hf_2InB_2$ | 1.16 | 0.67 | 0.78 | 1.25 | 499 | 400 | 0.177 |
| $Hf_2SnB_2$ | 1.04 | 0.76 | 0.79 | 1.12 | 514 | 460 | 0.040 |
| $Ti_2InB_2$ | 1.29 | 0.59 | 0.77 | 1.36 | 490 | 360 | 0.305 |

For isotropic solids, the values of $A_i$ are equal to 1; $k_c/k_a$ is equal to 1; $B_a$ is equal to $B_c$; $A^U$ is equal to 0, while the departure from these values indicate the anisotropic nature. As evident from the Table 5; the titled borides are anisotropic in nature. The anisotropy indices of $Zr_2AB_2$ (A = In, Tl) are comparable with that of the $Hf_2AB_2$ (A = In, Sn) and $Ti_2InB_2$ (Table 4 and Table 5).

### 3.3 Electronic band structure and density of states

The electronic band structures of $Zr_2AB_2$ (A = In, Tl) along high symmetry directions in the first Brillouin zone are illustrated in Fig. 6. The band structure clearly demonstrates the metallic nature with considerable overlapping of valence and conduction bands like other MAX phases [22,28,72,73]. The band structures also reveal the direction dependent electrical conductivity from the difference in the degree of energy dispersion. The band paths along $c$-direction: Γ-A, H-K and M-L exhibit lower degree of energy dispersion compared to the other paths that are in the basal planes: A-H, K-Γ, Γ-M and L-H [74]. The low level of energy dispersion confirms the higher electronic effective mass tensor for the conduction along *c*-axis in comparison with that in the *ab*-plane, consequently, lower electrical conductivity along the *c*-direction compared to the *ab*-basal plane should result. Thus, anisotropic electrical conductivity is predicted.

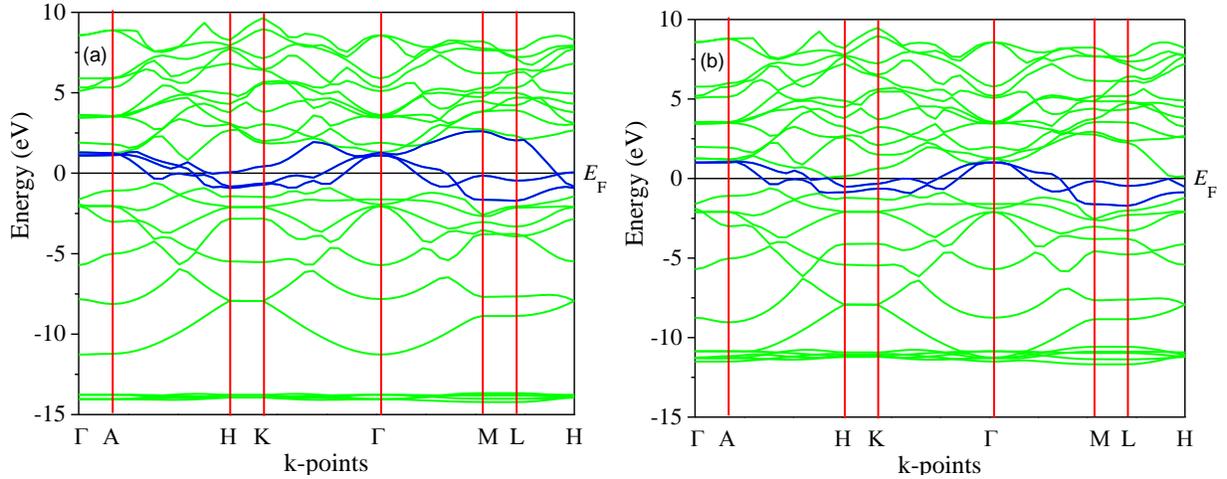

Fig. 6: The electronic band structure of (a) $Zr_2InB_2$ and (b) $Zr_2TlB_2$ compounds.

Furthermore, in order to explore the bonding nature from a different perspective, we have calculated the total and partial density of states (DOS) in addition to Mulliken bond population analysis. Fig. 7 demonstrated the total and partial DOS of $Zr_2AB_2$ (A = In, Tl) in which two red lines in the bottom panels are used to distinguish the energy regions of the valence band. The peaks in the lowest valence band come from the hybridization of B-2*s* and B-2*p* electronic states. The peaks in the middle valence band regions come from the hybridization of B-2*s*, B-2*p* and In-5s electronic states with a dominant contribution from the B-2*p* states. The upper valence band originates from the hybridization among Zr-4*d*, In-5*p*/Tl-5*p* and B-2*p* orbitals where Zr-4*d* states dominate the energy range. As seen from these two figures the pattern of the peaks in the DOS are almost identical for both compounds. The position of the peaks in DOS plays important role in the bonding strength between the hybridizing electronic states; the position in the low energy side indicates stronger bonding and vice versa. The position of the peaks in the DOS is also identical for both compounds like CDM results, thus, almost similar bonding strength (i.e. hardness) is expected for both compounds. But, Table 3 exhibits different Vickers hardness for these two compounds. Now, the question is what leads to this difference? To clarify this point, let us have a closer look at Table 3 again. The strongest bond within these compounds is the B-B bonding. The bond length of B-B bonding is smaller for $Zr_2InB_2$ (1.868 Å) compared to $Zr_2TlB_2$ (1.879 Å) that leads higher BOP for $Zr_2InB_2$ (2.22) compared to $Zr_2TlB_2$ (2.17). Thus, the bonding strength of B-B is higher in $Zr_2InB_2$ than that in $Zr_2TlB_2$. Similar variation in the bonding strength between Zr-B for $Zr_2InB_2$ and $Zr_2TlB_2$ is also noted in Table 3. The difference in the strengths and bond lengths of B-B and Zr-B bonds are therefore responsible for the variation in the elastic constants, elastic moduli and Vickers hardness between $Zr_2AB_2$ (A = In, Tl) as shown in Table 2 and Table 3.

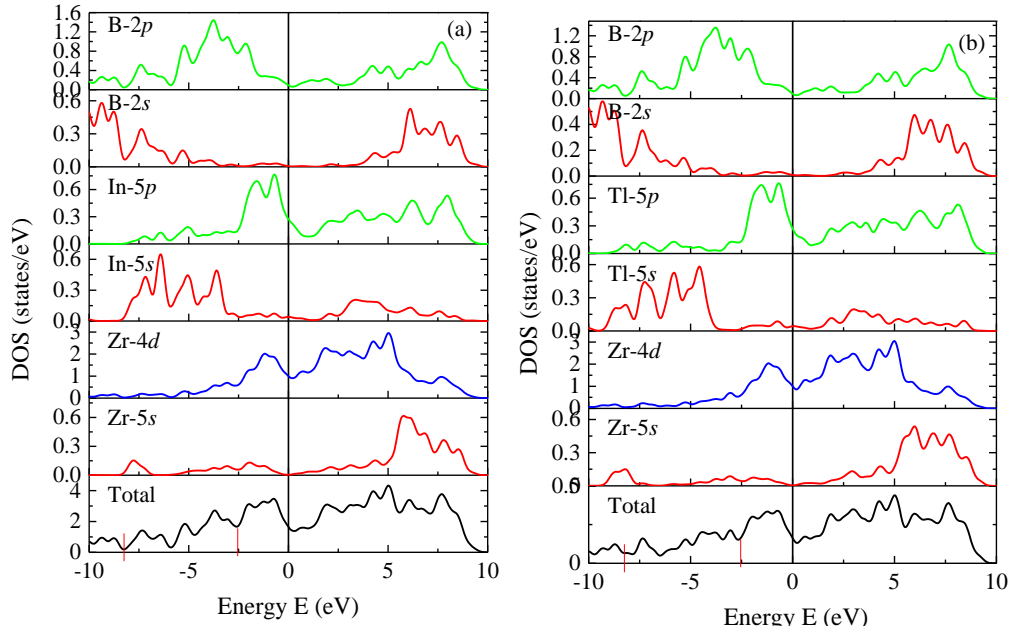

Fig. 7: Total and partial DOS of (a) $Zr_2InB_2$ and (b) $Zr_2TlB_2$ compounds.

### 3.4 Thermal Properties

Someimportant parameters that are used to characterize the thermal properties of materials to predict possible applications are calculated for the titled compounds and compared with those of similar systems in this section.

The collective thermal lattice vibration contributes to the specific heat in a solid. The phonon contribution to the specific heat is largely controlled by the Debye temperature $\Theta_D$. Various other thermodynamic properties such as thermal expansion, thermal conductivity, and lattice enthalpy are related to elastic properties and the $\Theta_D$. In general, high $\Theta_D$ value indicates enhanced hardness of the material. $\Theta_D$ can be calculated using the following equation [65]:

$$\Theta_D = h/k_B \left[(3n/4\pi)N_A\rho/M\right]^{1/3} v_m,$$

where, $M$, $n$, $\rho$, $h$, $k_B$, $N_A$ and $v_m$ are the molar mass, the number of atoms in the molecule, the mass density, the Planck's constant, Boltzmann constant, Avogadro's number, and average sound velocity ($v_m$) that is associated to shear modulus ($G$) and bulk modulus ($B$). The mean sound velocity $v_m$ which is expressed by the following equation:

$$v_m = \left[1/3\left(1/v_l^3 + 2/v_t^3\right)\right]^{-1/3}$$

where, $v_l$ and $v_t$ are the longitudinal and transverse sound velocities, respectively. $v_l$ and $v_t$ are represented by the following equations which are related to the elastic moduli (shear and bulk modulus) and density of the solid:

$$v_l = [(3B + 4G)/3\rho]^{1/2} \text{ and } v_t = [G/\rho]^{1/2}.$$

The Debye temperature, $\Theta_D$, the density ($\rho$), different sound velocities ($v_l$, $v_t$ and $v_m$) for studied compositions $Zr_2AB_2$ (A = In, Tl) are calculated and tabulated in Table 6 along with those of $Zr_2AC$ (A = In, Tl), $Hf_2AB_2$ (A = In, Sn), $Hf_2AC$ (A = In, Sn), $Ti_2InB_2$ and $Ti_2InC$ for comparison. The calculated values of $\Theta_D$ for $Zr_2AB_2$ (A = In, Tl) are higher than those of $Zr_2AC$ (A = In, Tl). In fact, the $\Theta_D$ of 212 MAX borides are higher than that of the 211 MAX phase carbides, is associated with the statement that $\Theta_D$ is higher for harder solids [75]. The compounds presented in Table 6 can be ranked based on the values of $\Theta_D$ as follows: $Ti_2InB_2 > Ti_2InC > Zr_2InB_2 > Zr_2InC > Hf_2SnB_2 > Zr_2TlB_2 > Hf_2InB_2 > Hf_2SnC > Hf_2InC > Zr_2TlC$. Higher value of $\Theta_D$ is desirable for the use of solids in high temperature technology; thus, 212 MAX phase borides exhibit better suitability compared to their 211 MAX phase carbide counterparts to be used for the same.

The thermal conductivity is the physical property of a material that measures its ability to conduct heat. It is closely associated with the acoustic wave velocity of phonons and phonon density of states inside the material. It decreases with increasing temperature and becomes saturated to a constant (minimum) value at very high temperature [76]. Hence, determination of minimum value of thermal conductivity is of importance in the application of materials in high-temperature conditions. The minimum thermal conductivity of the polycrystalline solids can be calculated based on the following equation due to Clark [76]:

$$K_{min} = k_B v_m \left(\frac{M}{n\rho N_A}\right)^{-\frac{2}{3}}$$

Here the symbols carry the same meanings as used in the equation of $\Theta_D$. As seen in Table 6, the values of $K_{min}$ for $Zr_2AB_2$ (A = In, Tl) are lower than that of $Zr_2AC_2$ (A = In, Tl) carbides, indicating their lower ability to conduct heat at high temperature. In case of thermal barrier coating (TBC) materials, $K_{min}$ should be low with some other desirable thermal parameters such as low thermal expansion coefficient (*TEC*), high melting point, etc. Thus, $Zr_2AB_2$ (A = In, Tl) borides are better candidate materials than $Zr_2AC$ (A = In, Tl) carbides for TBC purpose. In fact, $K_{min}$ of $Zr_2AB_2$ (A = In, Tl) borides is less than the value of 1.13 W/mK of the well-known TBC material $Y_4Al_2O_9$ [77,78]. The $K_{min}$ of $Hf_2AB_2$ (A = In, Sn) borides are also lower than those of $Hf_2AC$ (A = In, Sn) carbides. However, $Zr_2InC$ has the highest minimum thermal conductivity and $Zr_2TlB_2$ and $Hf_2InB_2$ have the lowest minimum thermal conductivity among the compounds presented in Table 6. Besides, the 212 MAX phase borides can be ranked based on the low value of $K_{min}$ as follows: $Zr_2TlB_2$>$Hf_2InB_2$>$Hf_2SnB_2$>$Zr_2TlC$ >$Zr_2InB_2$>$Ti_2InC$>$Hf_2InC$ > $Hf_2SnC$ >$Ti_2InB_2$>$Zr_2InC$.

**Table 6:** Calculated crystal density, longitudinal, transverse and average sound velocities ($v_l$, $v_t$, and $v_m$, respectively), Debye temperature, $\Theta_D$, minimum thermal conductivity, $K_{min}$, Grüneisen parameter, $\gamma$, melting temperature, $T_m$, and thermal expansion coefficient (*TEC*) of $Zr_2AB_2$ (A = In, Tl), together with those of $Hf_2AB_2$ (A = In, Sn), $Hf_2AC$ (A = In, Sn), $Ti_2InB_2$ and $Ti_2InC$ MAX phases.

| Phase | $\rho$ (g/cm$^3$) | $v_l$ (m/s) | $v_t$ (m/s) | $v_m$ (m/s) | $\Theta_D$ (K) | $K_{min}$ (W/mK) | $\gamma$ | $T_m$ (K) | *TEC* (10$^{-6}$ K$^{-1}$) | Ref. |
|---|---|---|---|---|---|---|---|---|---|---|
| $Zr_2InB_2$ | 6.87 | 6358 | 3907 | 4312 | 516 | 0.92 | 1.28 | 1693 | 6.5 | This study |
| $Zr_2InC$ | 7.32 | 6004 | 3616 | 3999 | 459 | 1.24 | 1.36 | 1584 | | Ref- [54] |
| $Zr_2TlB_2$ | 8.71 | 5466 | 3284 | 3633 | 433 | 0.77 | 1.36 | 1660 | 7.1 | This study |
| $Zr_2TlC$ | 8.92 | 4989 | 2993 | 3311 | 372 | 0.89 | 1.36 | 1430 | | Ref- [55] |
| $Hf_2InB_2$ | 10.86 | 5309 | 3240 | 3578 | 431 | 0.77 | 1.29 | 1800 | 5.9 | Ref- [28] |
| $Hf_2InC$ | 11.67 | 5004 | 2999 | 3319 | 383 | 1.04 | 1.32 | 1691 | | Ref- [54] |
| $Hf_2SnB_2$ | 11.08 | 5459 | 3344 | 3692 | 447 | 0.80 | 1.28 | 1872 | 5.6 | Ref- [28] |
| $Hf_2SnC$ | 12.06 | 5121 | 3050 | 3376 | 393 | 1.07 | 1.49 | 1746 | | Ref- [92] |
| $Ti_2InB_2$ | 05.90 | 7241 | 4545 | 5004 | 633 | 1.19 | 1.19 | 1858 | | Ref- [28] |
| | 05.91 | 7168 | 4487 | 4942 | 621 | 1.23 | 1.20 | 1833 | 15.4 | Ref- [36] |
| $Ti_2InC$ | 06.08 | 6531 | 4055 | 4471 | 534 | 1.00 | 1.23 | 1569 | | Ref- [36] |

We have also calculated the Grüneisen parameter ($\gamma$) that is used to quantify the anharmonic effects within the crystals. Study of the anharmonic effect is essential as it relates the specific heat at constant volume, bulk modulus, *TEC*, and volume together. $\gamma$ is computed from the Poisson's ratio using the equation [79]: $\gamma = \frac{3}{2}\frac{(1+\nu)}{(2-3\nu)}$. The calculated values of $\gamma$ as presented in the Table 6 indicating a low anharmonic effect within the compounds investigated herein. Moreover, the values are also within the limit of 0.85 to 3.53 for the solids with Poisson's ratio in the range of 0.05–0.46 [80].

The melting point ($T_m$) of solids gives the idea about the limit of temperature up to which the solid can be used. Thus, $T_m$ is critical for the selection materials to be used in any sector or devices to be used at high temperature. It is well-known that many of the MAX phase compounds are suitable for use in high temperature technology. From this point of view, we have calculated the melting temperatures of the titled borides using the equation based on the elastic constants as follows[81]: $T_m = 3C_{11}+1.5C_{33}+354$ and presented those in Table 6. The $T_m$ values of $Zr_2AB_2$ (A = In, Tl) are higher than those of $Zr_2AC$ (A = In, Tl). In fact, the $T_m$ of 212 MAX phase borides are higher than those of corresponding 211 MAX phase carbides (Table 6). The *Y* of $Zr_2AB_2$(A = In, Tl) is noticed to be 8.65% and 17.4% higher than that of $Zr_2AC$ (A = In, Tl), respectively, and the $T_m$ of $Zr_2AB_2$(A = In, Tl) is 6.9% and 16.1% higher than that of $Zr_2AC$ (A = In, Tl), respectively. For other 212 MAX phase borides, $T_m$ is significantly higher than those of 211 carbides presented here. Melting temperature provides information about bonding strength and is closely related with Young's modulus. Usually, a solid with higher Young's modulus has higher melting point and vice versa [82]. Thus, a significant increase in bonding strength for 212 MAX phase borides due to 2D B-layer is assumed to be responsible for the increase of $T_m$ compared to their carbide counterparts. A similar relationship is also observed for $Hf_3PB_4$ ($T_m$~2282 K) [37] for which the highest melting point is reported among the known MAX phases with the highest value of Young's modulus (426 GPa). Though the decomposition temperature ($T_d$) is not known but the $T_m$ indicates that the titled borides could be used in high-temperature technology within its decomposition limit. Cue *et al.*[83] have measured the $T_d$ of some MAX phases and the $T_m$ of those MAX phases have been reported [37]. The report showed that the $T_d$ is generally lower than $T_m$; for some cases $T_d$ is close to $T_m$. Thus, it is reasonable to expect that the decomposition temperature of the titled borides will also be high and they can be used up to that temperature.

The phonon specific heat capacity at constant volume ($C_v$) of the compounds can be calculated using following standard expression [84,85]:

$$C_v = 9nN_Ak_B \left(\frac{T}{\Theta_D}\right)\int_0^{x_D} dx \frac{x^4}{(e^x-1)^2}$$

where, $x_D = \frac{\Theta_D}{T}$ ; and $n$, $N_A$ and $k_B$ are the number of atoms per formula unit, the Avogadro's number and the Boltzmann constant, respectively. The linear thermal expansion coefficient ($\alpha$) and specific heat at constant pressure ($C_p$) are also calculated using following relations [76]:

$$\alpha = \frac{\gamma C_v}{3 B_T v_m} \text{ and } C_p = C_v(1 + \alpha \gamma T)$$

where, $B_T, v_m$ and $\gamma = 3(1+v)/[2(2-3v)]$ are the isothermal bulk modulus, molar volume and Grüneisen parameter, respectively.

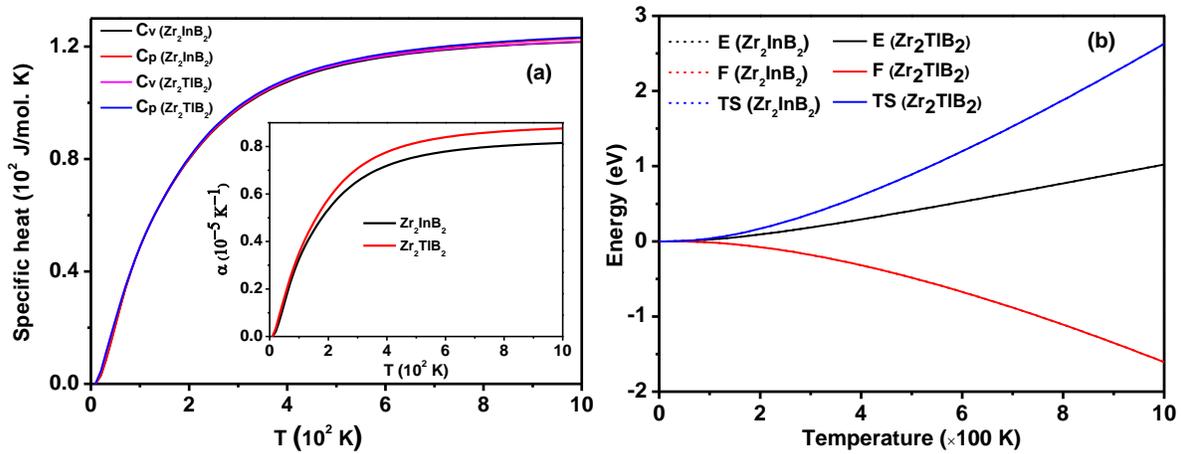

Fig. 8: (a) Temperature dependence of specific heat $C_v$ and $C_p$ for $Zr_2AB_2$ (A = In, Tl). The inset shows thermal expansion coefficient, $\alpha$ for the same. (b) Temperature dependence of the thermodynamic potential functions of $Zr_2AB_2$ (A = In, Tl) compounds.

The temperature dependence of specific heats, $C_v$, $C_p$, and $\alpha$ for the $Zr_2AB_2$ (A = In, Tl) compounds have been calculated using the formulae mentioned above and illustrated in Fig. 8 (a). These parameters are calculated in the temperature range of 0-1000 K where quasi-harmonic model is assumed to be valid and no phase transitions are expected for the compounds considered here. The heat capacity increases with the increase in temperature as the phonon thermal softening happens with increasing temperature. The heat capacities rise quickly at lower temperature regime and follow the Debye-$T^3$ power-law [86] and almost approaches to Dulong-Petit ($3nN_A k_B$) limit at higher temperature regime where the $C_v$ and $C_v$ do not depend strongly on the temperature[87]. The thermal expansion coefficient (*TEC*) of materials arise from anharmonicity in the lattice dynamics which is linked to the difference between the specific heats $C_p$ and $C_v$. The value of *TEC* for $Zr_2AB_2$ (A = In, Tl) compounds are estimated and shown in inset of Fig. 8(a). It is seen that *TEC* increases rapidly up to temperature 400 K and then gradually increases in the temperature ranging from 400 K to 700 K and finally approaches to saturate. The values of *TEC* are found to be $6.5 \times 10^{-6}$ and $7.1 \times 10^{-6}$, for the $Zr_2InB_2$ and $Zr_2TlB_2$ at temperature 300 K, respectively.

Quasi-harmonic approximation at zero pressure is used to calculate thermodynamic potential functions such as Helmholtz free energy *F*, internal energy *E*, entropy *S* for $Zr_2AB_2$ (A = In, Tl) compounds and are shown in Fig. 8 (b) [47]. It can be seen from Fig. 8 (b), below 100 K, the values of *E*, *F* and *TS* are almost zero. A nonlinear decrease (increase) of *F*(TS) is observed above 100 K. The decreasing trend of *F* is very common and it becomes more negative during the progress of any natural process. The *F* is defined as the difference between internal energy of a system and the amount of unusable energy to perform work. This unusable energy can be articulated as the product of S and the absolute temperature of the system. In order to compare the *F* and *E* we represent the entropy as *TS* in Fig. 8 (b). The thermal disturbance increases the disorder with increasing temperature in the system and as a consequence the entropy (S) increases as shown in Fig. 8 (b). An increasing trend of *E* with the increase of temperature is observed as expected for compounds of any type.

In recent times, the MAX phases have been considered as promising TBC materials [56,88] where the values of Debye temperature, minimum thermal conductivity, thermal expansion coefficient, melting temperature, etc. were taken into account. The widely considered TBC material, $Y_4Al_2O_9$, has the values of $\Theta_D \sim 564$ K, $K_{min} \sim 1.13$ W/mK, *TEC* $\sim 7.51 \times 10^{-6}$ K$^{-1}$, and $T_m \sim 2020$ K [77,78]. A comparison of the values of $\Theta_D$, $K_{min}$, *TEC*, and $T_m$ for $Zr_2AB_2$ (A = In, Tl) presented in Table 6 with those of the $Y_4Al_2O_9$ indicate that the $Zr_2AB_2$ (A = In, Tl) borides have the potential to be used as efficient TBC materials.

### *3.5 Optical properties*

The interactions between solids and electromagnetic radiation (photon) are of decisive importance for optoelectronic, photovoltaic applications as well as for analyzing the materials fundamental optoelectronic properties. The frequency/energy dependent dielectric function and the electronic band structure as well as the underlying DOS are closely related to each other. Since the electronic structure of the MAX compounds ensures the metallic conductivity, some additional parameters (plasma frequency, 3 eV; damping, 0.05 eV and Gaussian smearing, 0.5 eV) have been selected for analyzing the optical properties [89]. The imaginary part of dielectric function $\varepsilon_2(\omega)$ from momentum matrix element between the occupied and unoccupied electronic states is calculated using the following equation:

$$\varepsilon_2(\omega) = \frac{2e^2\pi}{\Omega \epsilon_0} \sum_{k,v,c} |\psi_k^c|u.r|\psi_k^v|^2 \delta(E_k^c - E_k^v - E)$$

where, $e$ = electronic charge; $\omega$ = light angular frequency; $u$ = polarization vector of incident electric field; $\psi_k^c$ = conduction band wave function at $k$ and $\psi_k^v$ = valence band wave function at $k$. Then the real part of dielectric function $\varepsilon_1(\omega)$ can be estimated from the well known Kramers-Kronig equation. All the other optical properties (absorption coefficient, photoconductivity, reflectivity and loss function) are evaluated using $\varepsilon_1(\omega)$ and $\varepsilon_2(\omega)$ [90]. In this section, various energy dependent optical properties of $Zr_2AB_2$ (A = In, Tl) 212 MAX phases are calculated and analyzed in detail for the photon energy range of 0 - 25 eV in order to assess the compounds for practical applications for the first time. Here, we have plotted relevant results of some other 212 MAX compounds $Ti_2InB_2$, $Hf_2InB_2$ and $Hf_2SnB_2$ with the studied compounds for comparison only. Noted here that the spectral behavior of $Ti_2InB_2$, $Hf_2InB_2$ and $Hf_2SnB_2$ MAX compounds are almost similar in the IR and visible energy rangeand only some changes of peak positions and intensities are observed in the UV region. It should be also noted here that the spectra presented here are only for the [100] polarization direction of the incident electric field, owing to the very low level of optical anisotropy reported for other MAX phase nanolaminates [8,27,28,66].

The dielectric function spectra have two components; the real part $\varepsilon_1(\omega)$ and the imaginary part $\varepsilon_2(\omega)$ where $\varepsilon_1(\omega)$ is linked to the polarizability and $\varepsilon_2(\omega)$ is associated with dielectric losses as a function of frequency. The energy dependent $\varepsilon_1(\omega)$ value which started from below, approached zero at ~16 eV, while the $\varepsilon_2(\omega)$ approached zero from above at ~14 eV as depicted in Fig. 9 (a-b). This phenomenon gives the confirmation of Drude like behavior (metallic feature) of the compounds under study. The spectra of $\varepsilon_1(\omega)$ in the low energy region (< 1 eV) results in the highest value of dielectric constant, is due to photon induced intraband transition of electrons. The spectra of $\varepsilon_2(\omega)$ was gradually decreased with energy with some prominent peaks in the infrared region (IR < 1.7 eV). The spectrums of dielectric function of other MAX materials ($Ti_2InB_2$, $Hf_2InB_2$ and $Hf_2SnB_2$) reveal almost similar characteristics.

The energy loss function (measure of the energy absorbed in the medium) is evaluated from the loss spectrum which also corresponds to the bulk plasma frequency ($\omega_p$) and ascends when $\varepsilon_2(\omega) < 1$ and $\varepsilon_1(\omega) = 0$. Prominent energy loss of a electron when travelling through the material with the highest velocity is noticed in the mid UV region (at around 16.5 eV for both compounds) as shown in Fig. 9 (c). The bulk plasma frequencies for $Ti_2InB_2$, $Hf_2InB_2$ and $Hf_2SnB_2$ were gradually shifted in the high energy region of electromagnetic radiation in the sequence of 17.4, 18.5 and 18.7 eV, respectively.

It is seen from Fig. 9 (d) that absorption spectra start rising at very low photon energy which is a manifestation of metallic behavior of the material. The absorption in the IR region is not significant, it increases in the visible light region vigorously and the maximum occurs in the UV region at ~11 eV and then decreases gradually, and finally no absorption is observed above 23 eV ($Zr_2TlB_2$) and 21 eV ($Zr_2TlB_2$) of photon energy. However, large absorption band mainly in the visible and ultraviolet region manifests that the titled materials can be used to design optoelectronic devices and even has strong possibility to be used in UV surface-disinfection device, medical sterilizer equipment etc. Photoconductivity behavior of these materials directly follows the absorption spectra of the materials. The highest conductivity was revealed for both compounds at ~3.1 eV energy and then decreased gradually with some prominent peaks. All these results clearly suggest that the studied materials should not have the band gap, which is also confirmed from electronic structure analysis.

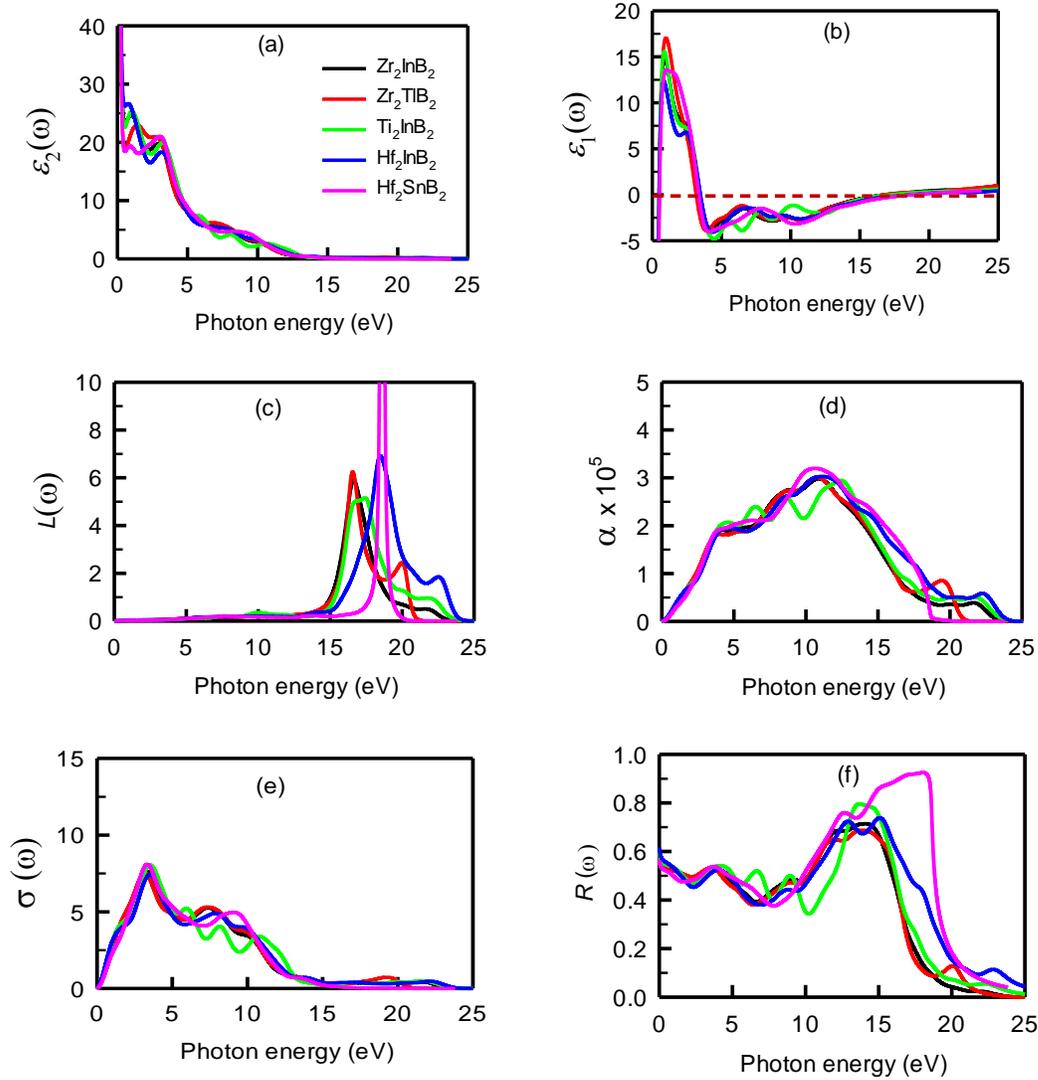

Fig. 9: The (a) real and (b) imaginary parts of dielectric function, (c) the energy loss function, (d) absorption coefficient, (e) photoconductivity and reflectivity of $Zr_2AB_2$ (A = In, Tl), together with results for $Zr_2AB_2$ (A = In, Sn) and $Ti_2InB_2$ as function of photon energy. The color scheme for different MAX phases is given in the first panel (a).

Reflectivity measures the fraction of the incident light energy reflected from the material. The reflectivity spectra started with value of ~0.6 (60%) for both the compounds as shown in Fig. 9 (f). The spectra have values always above 47% in the IR and visible region and the materials under study should be found as metallic gray in color. The highest reflectivity values of 71% ($Zr_2InB_2$) and 68% ($Zr_2TlB_2$) are found in the mid UV region at around 14.5 eV whereas those of 77% ($Ti_2InB_2$), 73% ($Hf_2InB_2$) and 92% ($Hf_2SnB_2$) are found at 14.50, 14.60 and 18.20 eV, respectively. The reflectivity of both the studied compounds drastically falls around photon energy ~15 eV. It has been reported that compounds with average value of reflectivity above 44% in the visible light region can be used to reduce solar heating by reflecting large portion of the solar spectrum [89,91]. Our result reveals that both the studied compounds should be capable to reduce solar heating.

## 4. Conclusions

This paper is focused on the investigation of physical properties of recently predicted $Zr_2AB_2$ (A = In, Tl) MAX phase borides using DFT calculations. The obtained lattice constants, crystal cell volume and atomic positions are consistent with the previous report [30]. The results of stiffness constants and phonon dispersion curves predicted their dynamical and mechanical stability. The mechanical properties characterizing parameters like stiffness constants, elastic moduli, machinability index, fracture mode, Cauchy pressure and hardness parameters are calculated and analyzed. The obtained values (mechanical properties) for $Zr_2AB_2$ (A = In, Tl) are higher than those of $Zr_2AC$ (A = In, Tl) owing to the existence of 2D boron layer within $Zr_2AB_2$ (A = In, Tl) compounds. The mechanical parameters of $Zr_2InB_2$ are better than those of $Zr_2TlB_2$ which have been explained meticulously based on the bonding strength, bond length and bond population analysis of B-B and Zr-B atoms. Moreover, $Zr_2AB_2$ (A = In, Tl) MAX phase boridesare members of brittle class of family with highly anisotropic elastic behaviors. The electronic band structure and density of states (DOS) confirm metallic character. Different chemical bonding naturehas been explored from the study of partial DOS and charge density mapping. The electrical conductivity at the Fermi level is mainly contributed by the Zr-*d* orbitals with a small contribution from the B/C-*p* orbitals. The anisotropy in electrical conductivity is also prominently noted from the energy dispersion curves along *c*-axis and within the *ab*-plane. Technologically important thermal parameters of $Zr_2AB_2$ (A = In, Tl) such as Debye temperature, minimum thermal conductivity, Grüneisen parameter and melting temperature are found to be higher than those of conventional $Zr_2AC$ (A = In, Tl) MAX compounds. The factors responsible for enhanced mechanical properties are also assumed to be responsible for this difference in the thermal properties because of their close relationship with the mechanical properties. The heat capacities ($C_v$ and $C_p$) and thermal expansion coefficient rise quickly in the temperature range 0 to 300 K and then attains a plateau at higher temperatures. The $C_v$ and $C_p$ obeyed Debye $T^3$-law at low temperature and attained classical Dulong-Petit (DP) limit at high temperature. The thermodynamic potential functions, such as Helmholtz free energy, internal energy and entropy show standard temperature dependence as expected for solids. The value of $K_{min}$, *TEC*, and $T_m$ suggest the possible use of $Zr_2AB_2$ (A = In, Tl) as thermal barrier coating materials for high-temperature devices. The optical constants spectra confirm the metallic nature of $Zr_2AB_2$ (A = In, Tl). The absorption spectra reveal the possibility of $Zr_2AB_2$ (A = In, Tl) to be used in optoelectronic devices, UV surface-disinfection device, medical sterilizer equipments etc. Furthermore, the reflectivity is high (> 44 %) in IR regions, visible and mid UV regions, suggesting their suitability as coating material to avoid solar heating.

**Acknowledgment**

The authors are grateful to the Department of Physics, Chittagong University of Engineering & Technology (CUET), Chattogram-4349, Bangladesh, for providing the computing facility for this work.

**References**

[1] M.W. Barsoum, G. Yaroschuk, S. Tyagi, Fabrication and characterization of M2SnC (M = Ti, Zr, Hf and Nb), Scr. Mater. 37 (1997) 1583–1591. https://doi.org/10.1016/S1359-6462(97)00288-1.

[2] M.W. Barsoum, T. El-Raghy, Synthesis and Characterization of a Remarkable Ceramic: Ti3SiC2, J. Am. Ceram. Soc. 79 (1996) 1953–1956. https://doi.org/10.1111/j.1151-2916.1996.tb08018.x.

[3] M.W. Barsoum, A progress report on Ti3SiC2, Ti3GeC2, and the H-phases, M2BX, J. Mater. Synth. Process. (1997).

[4] M.R. Khatun, M.A. Ali, F. Parvin, A.K.M.A. Islam, Elastic, thermodynamic and optical behavior of V 2 A C ( A = Al, Ga) MAX phases, Results Phys. 7 (2017) 3634–3639. https://doi.org/10.1016/j.rinp.2017.09.043.

[5] M.A. Ali, M.T. Nasir, M.R. Khatun, A.K.M.A. Islam, S.H. Naqib, An *ab initio* investigation of vibrational, thermodynamic, and optical properties of Sc 2 AlC MAX compound, Chinese Phys. B. 25 (2016) 103102. https://doi.org/10.1088/1674-1056/25/10/103102.

[6] M.A. Ali, M.R. Khatun, N. Jahan, M.M. Hossain, Comparative study of Mo 2 Ga 2 C with superconducting *MAX* phase Mo 2 GaC: First-principles calculations, Chinese Phys. B. 26 (2017) 033102. https://doi.org/10.1088/1674-1056/26/3/033102.

[7] F. Sultana, M.M. Uddin, M.A. Ali, M.M. Hossain, S.H. Naqib, A.K.M.A. Islam, First principles study of M2InC (M = Zr, Hf and Ta) MAX phases: The effect of M atomic species, Results Phys. 11 (2018) 869–876. https://doi.org/10.1016/j.rinp.2018.10.044.

[8] A. Chowdhury, M.A. Ali, M.M. Hossain, M.M. Uddin, S.H. Naqib, A.K.M.A. Islam, Predicted MAX Phase Sc 2 InC: Dynamical Stability, Vibrational and Optical Properties, Phys. Status Solidi. 255 (2017) 1700235. https://doi.org/10.1002/pssb.201700235.

[9] C. Zuo, C. Zhong, Screen the elastic and thermodynamic properties of MAX solid

solution using DFT procedue: Case study on (Ti1-xVx)2AlC, Mater. Chem. Phys. 250 (2020) 123059. https://doi.org/10.1016/j.matchemphys.2020.123059.

[10] Y.X. Wang, Z.X. Yan, W. Liu, G.L. Zhou, Structure stability, mechanical properties and thermal conductivity of the new hexagonal ternary phase Ti2InB2 under pressure, Philos. Mag. 100 (2020) 2054–2067. https://doi.org/10.1080/14786435.2020.1754485.

[11] P. Chakraborty, A. Chakrabarty, A. Dutta, T. Saha-Dasgupta, Soft MAX phases with boron substitution: A computational prediction, Phys. Rev. Mater. 2 (2018) 103605. https://doi.org/10.1103/PhysRevMaterials.2.103605.

[12] M.W. Barsoum, The MN+1AXN phases: A new class of solids, Prog. Solid State Chem. 28 (2000) 201–281. https://doi.org/10.1016/S0079-6786(00)00006-6.

[13] M.W. Barsoum, MAX phases: Properties of machinable ternary carbides and nitrides, Wiley-VCH Verlag GmbH & Co. KGaA, Weinheim, Germany, 2013. https://doi.org/10.1002/9783527654581.

[14] M. Radovic, M.W. Barsoum, MAX phases: Bridging the gap between metals and ceramics, Am. Ceram. Soc. Bull. 92 (2013) 20–27.

[15] A.S. Ingason, M. Dahlqvist, J. Rosen, Magnetic MAX phases from theory and experiments; A review, J. Phys. Condens. Matter. (2016). https://doi.org/10.1088/0953-8984/28/43/433003.

[16] M.A. Ali, Newly Synthesized Ta based MAX phase (Ta 1- x Hf x ) 4 AlC 3 and (Ta 1-x Hf x ) 4 Al 0.5 Sn 0.5 C 3 (0 ≤ x ≤ 0.25) Solid Solutions: Unravelling the Mechanical, Electronic and Thermodynamic Properties, Phys. Status Solidi. (2020) pssb.202000307. https://doi.org/10.1002/pssb.202000307.

[17] M. Naguib, V.N. Mochalin, M.W. Barsoum, Y. Gogotsi, 25th anniversary article: MXenes: A new family of two-dimensional materials, Adv. Mater. 26 (2014). https://doi.org/10.1002/adma.201304138.

[18] B. Anasori, M.R. Lukatskaya, Y. Gogotsi, 2D metal carbides and nitrides (MXenes) for energy storage, Nat. Rev. Mater. 2 (2017). https://doi.org/10.1038/natrevmats.2016.98.

[19] M. Naguib, M. Kurtoglu, V. Presser, J. Lu, J. Niu, M. Heon, L. Hultman, Y. Gogotsi, M.W. Barsoum, Two-Dimensional Nanocrystals: Two-Dimensional Nanocrystals Produced by Exfoliation of Ti3AlC2 (Adv. Mater. 37/2011), Adv. Mater. 23 (2011). https://doi.org/10.1002/adma.201190147.

[20] T. Rackl, L. Eisenburger, R. Niklaus, D. Johrendt, Syntheses and physical properties of the MAX phase boride Nb2SB and the solid solutions N b2 S Bx C1-x(x=0-1), Phys. Rev. Mater. 3 (2019) 054001. https://doi.org/10.1103/PhysRevMaterials.3.054001.

[21] T. Rackl, D. Johrendt, The MAX phase borides Zr2SB and Hf2SB, Solid State Sci. 106 (2020) 106316. https://doi.org/10.1016/j.solidstatesciences.2020.106316.

[22] M.A. Ali, M.M. Hossain, M.M. Uddin, M.A. Hossain, A.K.M.A. Islam, S.H. Naqib, Physical properties of new MAX phase borides M 2 SB (M = Zr, Hf and Nb) in comparison with conventional MAX phase carbides M 2 SC (M = Zr, Hf and Nb): Comprehensive insights, J. Mater. Res. Technol. 11 (2021) 1000–1018. https://doi.org/10.1016/j.jmrt.2021.01.068.

[23] C. Hu, C.-C. Lai, Q. Tao, J. Lu, J. Halim, L. Sun, J. Zhang, J. Yang, B. Anasori, J. Wang, Y. Sakka, L. Hultman, P. Eklund, J. Rosen, M.W. Barsoum, Mo 2 Ga 2 C: a new ternary nanolaminated carbide, Chem. Commun. 51 (2015) 6560–6563. https://doi.org/10.1039/C5CC00980D.

[24] H. Chen, D. Yang, Q. Zhang, S. Jin, L. Guo, J. Deng, X. Li, X. Chen, A Series of MAX

Phases with MA-Triangular-Prism Bilayers and Elastic Properties, Angew. Chemie - Int. Ed. 58 (2019) 4576–4580. https://doi.org/10.1002/anie.201814128.

[25] Q. Tao, J. Lu, M. Dahlqvist, A. Mockute, S. Calder, A. Petruhins, R. Meshkian, O. Rivin, D. Potashnikov, E.N. Caspi, H. Shaked, A. Hoser, C. Opagiste, R.-M. Galera, R. Salikhov, U. Wiedwald, C. Ritter, A.R. Wildes, B. Johansson, L. Hultman, M. Farle, M.W. Barsoum, J. Rosen, Atomically Layered and Ordered Rare-Earth i -MAX Phases: A New Class of Magnetic Quaternary Compounds, Chem. Mater. 31 (2019) 2476–2485. https://doi.org/10.1021/acs.chemmater.8b05298.

[26] L. Chen, M. Dahlqvist, T. Lapauw, B. Tunca, F. Wang, J. Lu, R. Meshkian, K. Lambrinou, B. Blanpain, J. Vleugels, J. Rosen, Theoretical Prediction and Synthesis of (Cr 2/3 Zr 1/3 ) 2 AlC i -MAX Phase, Inorg. Chem. 57 (2018) 6237–6244. https://doi.org/10.1021/acs.inorgchem.8b00021.

[27] M.A. Ali, M.M. Hossain, A.K.M.A. Islam, S.H. Naqib, Ternary boride Hf3PB4: Insights into the physical properties of the hardest possible boride MAX phase, J. Alloys Compd. 857 (2021) 158264. https://doi.org/10.1016/j.jallcom.2020.158264.

[28] M.A. Ali, M.M. Hossain, M.M. Uddin, A.K.M.A. Islam, D. Jana, S.H. Naqib, DFT insights into new B-containing 212 MAX phases: Hf2AB2 (A = In, Sn), J. Alloys Compd. 860 (2021) 158408. https://doi.org/10.1016/j.jallcom.2020.158408.

[29] J. Wang, T.N. Ye, Y. Gong, J. Wu, N. Miao, T. Tada, H. Hosono, Discovery of hexagonal ternary phase Ti2InB2 and its evolution to layered boride TiB, Nat. Commun. 10 (2019) 1–8. https://doi.org/10.1038/s41467-019-10297-8.

[30] N. Miao, J. Wang, Y. Gong, J. Wu, H. Niu, S. Wang, K. Li, A.R. Oganov, T. Tada, H. Hosono, Computational Prediction of Boron-Based MAX Phases and MXene Derivatives, Chem. Mater. 32 (2020) 6947–6957. https://doi.org/10.1021/acs.chemmater.0c02139.

[31] M. Ade, H. Hillebrecht, Ternary Borides Cr 2 AlB 2 , Cr 3 AlB 4 , and Cr 4 AlB 6 : The First Members of the Series (CrB 2 ) n CrAl with n = 1, 2, 3 and a Unifying Concept for Ternary Borides as MAB-Phases, Inorg. Chem. 54 (2015) 6122–6135. https://doi.org/10.1021/acs.inorgchem.5b00049.

[32] M.A. Ali, M.A. Hadi, M.M. Hossain, S.H. Naqib, A.K.M.A. Islam, Theoretical investigation of structural, elastic, and electronic properties of ternary boride MoAlB, Phys. Status Solidi. 254 (2017) 1700010. https://doi.org/10.1002/pssb.201700010.

[33] X. Li, H. Cui, R. Zhang, First-principles study of the electronic and optical properties of a new metallic MoAlB, Sci. Rep. 6 (2016) 39790. https://doi.org/10.1038/srep39790.

[34] S. Kota, M. Sokol, M.W. Barsoum, A progress report on the MAB phases: atomically laminated, ternary transition metal borides, Int. Mater. Rev. 65 (2020) 226–255. https://doi.org/10.1080/09506608.2019.1637090.

[35] Y.X. Wang, Z.X. Yan, W. Liu, G.L. Zhou, Structure stability, mechanical properties and thermal conductivity of the new hexagonal ternary phase Ti2InB2 under pressure, Philos. Mag. 100 (2020) 2054–2067. https://doi.org/10.1080/14786435.2020.1754485.

[36] M.M. Ali, M.A. Hadi, I. Ahmed, A.F.M.Y. Haider, A.K.M.. Islam, Physical properties of a novel boron-based ternary compound Ti2InB2, Mater. Today Commun. 25 (2020) 101600. https://doi.org/10.1016/j.mtcomm.2020.101600.

[37] M.A. Ali, M.M. Hossain, A.K.M.A. Islam, S.H. Naqib, Ternary boride Hf3PB4: Insights into the physical properties of the hardest possible boride MAX phase, J. Alloys Compd. 857 (2021) 158264. https://doi.org/10.1016/j.jallcom.2020.158264.

[38] A. Gencer, G. Surucu, Electronic and lattice dynamical properties of Ti 2 SiB MAX

phase, Mater. Res. Express. 5 (2018) 076303. https://doi.org/10.1088/2053-1591/aace7f.

[39] G. Surucu, A. Gencer, X. Wang, O. Surucu, Lattice dynamical and thermo-elastic properties of M2AlB (M = V, Nb, Ta) MAX phase borides, J. Alloys Compd. 819 (2020) 153256. https://doi.org/10.1016/j.jallcom.2019.153256.

[40] M. Khazaei, M. Arai, T. Sasaki, M. Estili, Y. Sakka, Trends in electronic structures and structural properties of MAX phases: a first-principles study on M 2 AlC (M = Sc, Ti, Cr, Zr, Nb, Mo, Hf, or Ta), M 2 AlN, and hypothetical M 2 AlB phases, J. Phys. Condens. Matter. 26 (2014) 505503. https://doi.org/10.1088/0953-8984/26/50/505503.

[41] G. Surucu, Investigation of structural, electronic, anisotropic elastic, and lattice dynamical properties of MAX phases borides: An Ab-initio study on hypothetical MAB (M = Ti, Zr, Hf; A = Al, Ga, In) compounds, Mater. Chem. Phys. 203 (2018) 106–117. https://doi.org/10.1016/j.matchemphys.2017.09.050.

[42] M.D. Segall, P.J.D. Lindan, M.J. Probert, C.J. Pickard, P.J. Hasnip, S.J. Clark, M.C. Payne, First-principles simulation: ideas, illustrations and the CASTEP code, J. Phys. Condens. Matter. 14 (2002) 2717–2744. https://doi.org/10.1088/0953-8984/14/11/301.

[43] S.J. Clark, M.D. Segall, C.J. Pickard, P.J. Hasnip, M.I.J. Probert, K. Refson, M.C. Payne, First principles methods using CASTEP, Zeitschrift Für Krist. - Cryst. Mater. 220 (2005). https://doi.org/10.1524/zkri.220.5.567.65075.

[44] J.P. Perdew, K. Burke, M. Ernzerhof, Generalized Gradient Approximation Made Simple, Phys. Rev. Lett. 77 (1996) 3865–3868. https://doi.org/10.1103/PhysRevLett.77.3865.

[45] H.J. Monkhorst, J.D. Pack, Special points for Brillouin-zone integrations, Phys. Rev. B. 13 (1976) 5188–5192. https://doi.org/10.1103/PhysRevB.13.5188.

[46] T.H. Fischer, J. Almlof, General methods for geometry and wave function optimization, J. Phys. Chem. 96 (1992) 9768–9774. https://doi.org/10.1021/j100203a036.

[47] S. Baroni, S. De Gironcoli, A. Dal Corso, P. Giannozzi, Phonons and related crystal properties from density-functional perturbation theory, Rev. Mod. Phys. 73 (2001) 515–562. https://doi.org/10.1103/RevModPhys.73.515.

[48] M.A. Ali, A.K.M.A. Islam, Sn1−xBixO2 and Sn1−xTaxO2 (0≤x≤0.75): A first-principles study, Phys. B Condens. Matter. 407 (2012) 1020–1026. https://doi.org/10.1016/j.physb.2012.01.002.

[49] M.A. Ali, A.K.M.A. Islam, N. Jahan, S. Karimunnesa, First-principles study of SnO under high pressure, Int. J. Mod. Phys. B. 30 (2016) 1650228. https://doi.org/10.1142/S0217979216502283.

[50] M.A. Ali, N. Jahan, A.K.M.A. Islam, Sulvanite Compounds $Cu_3TMS_4$ (TM = V, Nb and Ta): Elastic, Electronic, Optical and Thermal Properties using First-principles Method, J. Sci. Res. 6 (2014) 407–419. https://doi.org/10.3329/jsr.v6i3.19191.

[51] M.A. Ali, M. Roknuzzaman, M.T. Nasir, A.K.M.A. Islam, S.H. Naqib, Structural, elastic, electronic and optical properties of Cu3MTe4 (M = Nb, Ta) sulvanites — An *ab initio* study, Int. J. Mod. Phys. B. 30 (2016) 1650089. https://doi.org/10.1142/S0217979216500892.

[52] M. Born, On the stability of crystal lattices. I, Math. Proc. Cambridge Philos. Soc. 36 (1940) 160–172. https://doi.org/10.1017/S0305004100017138.

[53] F. Mouhat, F.-X. Coudert, Necessary and sufficient elastic stability conditions in various crystal systems, Phys. Rev. B. 90 (2014) 224104. https://doi.org/10.1103/PhysRevB.90.224104.

[54] A. Bouhemadou, Calculated structural and elastic properties of M2InC (M = Sc, Ti, V, Zr,

Nb, Hf, Ta), Mod. Phys. Lett. B. 22 (2008) 2063–2076. https://doi.org/10.1142/S0217984908016807.

[55] A. Bouhemadou, Structural, electronic and elastic properties of Ti2TlC, Zr2TlC and Hf2TlC, Cent. Eur. J. Phys. 7 (2009) 753–761. https://doi.org/10.2478/s11534-009-0022-z.

[56] M.A. Hadi, N. Kelaidis, S.H. Naqib, A. Chroneos, A.K.M.A. Islam, Mechanical behaviors, lattice thermal conductivity and vibrational properties of a new MAX phase Lu2SnC, J. Phys. Chem. Solids. 129 (2019) 162–171. https://doi.org/10.1016/j.jpcs.2019.01.009.

[57] R. Hill, The Elastic Behaviour of a Crystalline Aggregate, Proc. Phys. Soc. Sect. A. 65 (1952) 349–354. https://doi.org/10.1088/0370-1298/65/5/307.

[58] M.A. Ali, A.K.M.A. Islam, M.S. Ali, Ni-rich Nitrides ANNi$_3$ (A = Pt, Ag, Pd) in Comparison with Superconducting ZnNNi$_3$, J. Sci. Res. 4 (2011) 1. https://doi.org/10.3329/jsr.v4i1.9026.

[59] W. Voigt, Lehrbuch der Kristallphysik, Vieweg+Teubner Verlag, Wiesbaden, 1966. https://doi.org/10.1007/978-3-663-15884-4.

[60] A. Reuss, Berechnung der Fließgrenze von Mischkristallen auf Grund der Plastizitätsbedingung für Einkristalle ., ZAMM - J. Appl. Math. Mech. / Zeitschrift Für Angew. Math. Und Mech. 9 (1929) 49–58. https://doi.org/10.1002/zamm.19290090104.

[61] M.A. Ali, M. Anwar Hossain, M.A. Rayhan, M.M. Hossain, M.M. Uddin, M. Roknuzzaman, K. Ostrikov, A.K.M.A. Islam, S.H. Naqib, First-principles study of elastic, electronic, optical and thermoelectric properties of newly synthesized K2Cu2GeS4 chalcogenide, J. Alloys Compd. 781 (2018) 37–46. https://doi.org/10.1016/j.jallcom.2018.12.035.

[62] A. Bouhemadou, First-principles study of structural, electronic and elastic properties of Nb4AlC3, Brazilian J. Phys. 40 (2010) 52–57. https://doi.org/10.1590/S0103-97332010000100009.

[63] X.-Q. Chen, H. Niu, D. Li, Y. Li, Modeling hardness of polycrystalline materials and bulk metallic glasses, Intermetallics. 19 (2011) 1275–1281. https://doi.org/10.1016/j.intermet.2011.03.026.

[64] H. Gou, L. Hou, J. Zhang, F. Gao, Pressure-induced incompressibility of ReC and effect of metallic bonding on its hardness, Appl. Phys. Lett. 92 (2008) 241901. https://doi.org/10.1063/1.2938031.

[65] M.T. Nasir, M.A. Hadi, M.A. Rayhan, M.A. Ali, M.M. Hossain, M. Roknuzzaman, S.H. Naqib, A.K.M.A. Islam, M.M. Uddin, K. Ostrikov, First-Principles Study of Superconducting ScRhP and ScIrP pnictides, Phys. Status Solidi. 254 (2017) 1700336. https://doi.org/10.1002/pssb.201700336.

[66] M. Roknuzzaman, M.A. Hadi, M.A. Ali, M.M. Hossain, N. Jahan, M.M. Uddin, J.A. Alarco, K. Ostrikov, First hafnium-based MAX phase in the 312 family, Hf 3 AlC 2 : A first-principles study, J. Alloys Compd. 727 (2017) 616–626. https://doi.org/10.1016/j.jallcom.2017.08.151.

[67] M.A. Hadi, S.R.G. Christopoulos, S.H. Naqib, A. Chroneos, M.E. Fitzpatrick, A.K.M.A. Islam, Physical properties and defect processes of M3SnC2 (M = Ti, Zr, Hf) MAX phases: Effect of M-elements, J. Alloys Compd. 748 (2018) 804–813. https://doi.org/10.1016/j.jallcom.2018.03.182.

[68] D.G. Pettifor, Theoretical predictions of structure and related properties of intermetallics,

Mater. Sci. Technol. 8 (1992) 345–349. https://doi.org/10.1179/026708392790170801.
[69] H. Ledbetter, A. Migliori, A general elastic-anisotropy measure, J. Appl. Phys. 100 (2006) 063516. https://doi.org/10.1063/1.2338835.
[70] J. Chang, G.-P. Zhao, X.-L. Zhou, K. Liu, L.-Y. Lu, Structure and mechanical properties of tantalum mononitride under high pressure: A first-principles study, J. Appl. Phys. 112 (2012) 083519. https://doi.org/10.1063/1.4759279.
[71] R. Gaillac, P. Pullumbi, F.-X. Coudert, ELATE: an open-source online application for analysis and visualization of elastic tensors, J. Phys. Condens. Matter. 28 (2016) 275201. https://doi.org/10.1088/0953-8984/28/27/275201.
[72] M.A. Ali, M.M. Hossain, N. Jahan, A.K.M.A. Islam, S.H. Naqib, Newly synthesized $Zr_2AlC$, $Zr_2(Al_{0.58}Bi_{0.42})C$, $Zr_2(Al_{0.2}Sn_{0.8})C$, and $Zr_2(Al_{0.3}Sb_{0.7})C$ MAX phases: A DFT based first-principles study, Comput. Mater. Sci. 131 (2017) 139–145. https://doi.org/10.1016/j.commatsci.2017.01.048.
[73] M.A. Ali, S.H. Naqib, Recently synthesized $(Ti_{1-x}Mo_x)_2AlC$ ($0 \leq x \leq 0.20$) solid solutions: deciphering the structural, electronic, mechanical and thermodynamic properties via ab initio simulations, RSC Adv. 10 (2020) 31535–31546. https://doi.org/10.1039/D0RA06435A.
[74] Y. Zhou, Z. Sun, Electronic structure and bonding properties of layered machinable and ceramics, Phys. Rev. B - Condens. Matter Mater. Phys. 61 (2000) 12570–12573. https://doi.org/10.1103/PhysRevB.61.12570.
[75] M.A. Ali, M.M. Hossain, M.A. Hossain, M.T. Nasir, M.M. Uddin, M.Z. Hasan, A.K.M.A. Islam, S.H. Naqib, Recently synthesized $(Zr_{1-x}Ti_x)_2AlC$ ($0 \leq x \leq 1$) solid solutions: Theoretical study of the effects of M mixing on physical properties, J. Alloys Compd. 743 (2018) 146–154. https://doi.org/10.1016/j.jallcom.2018.01.396.
[76] D.R. Clarke, Materials selections guidelines for low thermal conductivity thermal barrier coatings, Surf. Coatings Technol. 163–164 (2003) 67–74. https://doi.org/10.1016/S0257-8972(02)00593-5.
[77] Y. Zhou, X. Lu, H. Xiang, Z. Feng, Preparation, mechanical, and thermal properties of a promising thermal barrier material: $Y_4Al_2O_9$, J. Adv. Ceram. 4 (2015) 94–102. https://doi.org/10.1007/s40145-015-0141-5.
[78] Y. Zhou, H. Xiang, X. Lu, Z. Feng, Z. Li, Theoretical prediction on mechanical and thermal properties of a promising thermal barrier material: $Y_4Al_2O_9$, J. Adv. Ceram. 4 (2015) 83–93. https://doi.org/10.1007/s40145-015-0140-6.
[79] V.N. Belomestnykh, E.P. Tesleva, Interrelation between anharmonicity and lateral strain in quasi-isotropic polycrystalline solids, Tech. Phys. 49 (2004) 1098–1100. https://doi.org/10.1134/1.1787679.
[80] S.I. Mikitishin, Interrelationship of Poisson's ratio with other characteristics of pure metals, Sov. Mater. Sci. 18 (1982) 262–265. https://doi.org/10.1007/BF01150837.
[81] M.E. Fine, L.D. Brown, H.L. Marcus, Elastic constants versus melting temperature in metals, Scr. Metall. 18 (1984) 951–956. https://doi.org/10.1016/0036-9748(84)90267-9.
[82] D. Salamon, Advanced Ceramics, in: Adv. Ceram. Dent., Elsevier Inc., 2014: pp. 103–122. https://doi.org/10.1016/B978-0-12-394619-5.00006-7.
[83] B. Cui, W.E. Lee, B. Cui, W.E. Lee, High-temperature Oxidation Behaviour of MAX Phase Ceramics, Refract. Worldforum. 5 (2013) 105–112.
[84] M.A. Blanco, E. Francisco, V. Luaña, GIBBS: Isothermal-isobaric thermodynamics of solids from energy curves using a quasi-harmonic Debye model, Comput. Phys. Commun.

158 (2004) 57–72. https://doi.org/10.1016/j.comphy.2003.12.001.

[85] O. Delaire, A.F. May, M.A. McGuire, W.D. Porter, M.S. Lucas, M.B. Stone, D.L. Abernathy, V.A. Ravi, S.A. Firdosy, G.J. Snyder, Phonon density of states and heat capacity of La3-x Te4, Phys. Rev. B - Condens. Matter Mater. Phys. 80 (2009) 184302. https://doi.org/10.1103/PhysRevB.80.184302.

[86] P. Debye, Zur Theorie der spezifischen Wärmen, Ann. Phys. 344 (1912) 789–839. https://doi.org/10.1002/andp.19123441404.

[87] A. Irvine, Alexis-Thérèse Petit ( 1791-1820 ) and Pierre-Louis Dulong, 413 (1838).

[88] M.A. Hadi, M. Dahlqvist, S.-R.G. Christopoulos, S.H. Naqib, A. Chroneos, A.K.M.A. Islam, Chemically stable new MAX phase V 2 SnC: a damage and radiation tolerant TBC material, RSC Adv. 10 (2020) 43783–43798. https://doi.org/10.1039/D0RA07730E.

[89] S. Li, R. Ahuja, M.W. Barsoum, P. Jena, B. Johansson, Optical properties of Ti3SiC2 and Ti4AlN3, Appl. Phys. Lett. 92 (2008) 221907. https://doi.org/10.1063/1.2938862.

[90] R. John, B. Merlin, Optical properties of graphene, silicene, germanene, and stanene from IR to far UV – A first principles study, J. Phys. Chem. Solids. 110 (2017) 307–315. https://doi.org/10.1016/j.jpcs.2017.06.026.

[91] M.A. Ali, M.S. Ali, M.M. Uddin, Structural, elastic, electronic and optical properties of metastable MAX phase Ti5SiC4 compound, Indian J. Pure Appl. Phys. 54 (2016) 386–390.

[92] A. Bouhemadou, Prediction study of structural and elastic properties under pressure effect of M2SnC (M=Ti, Zr, Nb, Hf), Phys. B Condens. Matter. 403 (2008) 2707–2713. https://doi.org/10.1016/j.physb.2008.02.014.